\documentclass{emulateapj}
\usepackage{natbib}
\usepackage{amssymb,amsmath} 

\shorttitle{Prompt and Reverse Shock Emission of GRB\,140512A}
\shortauthors{Huang et al. }
\begin{document}
\title{Very Bright Prompt and Reverse Shock Emission of GRB\,140512A}
\author{Xiao-Li Huang\altaffilmark{1,2}, Li-Ping Xin \altaffilmark{3}, Shuang-Xi Yi\altaffilmark{4}, Shu-Qing Zhong\altaffilmark{1,2}, Yu-Lei Qiu\altaffilmark{3}, Jin-Song Deng\altaffilmark{3}, Jian-Yan Wei\altaffilmark{3}, En-Wei Liang \altaffilmark{1,2,3}}

  \altaffiltext{1}{GXU-NAOC Center for Astrophysics and Space Sciences, Department of Physics, Guangxi University, Nanning 530004, China; lew@gxu.edu.cn}
  \altaffiltext{2}{Guangxi Key Laboratory for the Relativistic Astrophysics, Nanning 530004, China}
  \altaffiltext{3}{Key Laboratory of Space Astronomy and Technology, National Astronomical Observatories, Chinese Academy of Sciences, Beijing 100012, China; xlp@nao.cas.cn}
  \altaffiltext{4}{College of Physics and Engineering, Qufu Normal University, Qufu 273165, China}

\begin{abstract}
We report our observations of very bright prompt optical and reverse shock (RS) optical emission of GRB\,140512A and analyze its multi-wavelength data observed with the {\em Swift} and {\em Fermi} missions. It is found that the joint optical-X-ray-gamma-ray spectrum with our first optical detection (R=13.09 mag) at $T_0+136$ seconds during the second episode of the prompt gamma-rays can be fit by a single power-law with index $-1.32\pm 0.01$. Our empirical fit to the afterglow lightcurves indicates that the observed bright optical afterglow with R=13.00 mag at the peak time is consistent with predictions of the RS and forward shock (FS) emission of external shock models. Joint optical-X-ray afterglow spectrum is well fit with an absorbed single power-law, with an index evolving with time from $-1.86\pm 0.01$ at the peak time to $-1.57\pm 0.01$ at late epoch, which could be due to the evolution of the ratio of the RS to FS emission fluxes. We fit the lightcurves with standard external models, and derive the physical properties of the outflow. It is found that the ratio $R_B\equiv\epsilon_{\rm B, r}/\epsilon_{\rm B, f}$ is 8187, indicating a high magnetization degree in the RS region. Measuring the relative radiation efficiency with $R_e\equiv\epsilon_{\rm e, r}/\epsilon_{\rm e, f}$, we have $R_e= 0.02$, implying the radiation efficiency of the RS is much lower than that in FS. We also show that the $R_B$ of GRBs\,990123, 090102, and 130427A are similar to that of GRB\,140512A and their apparent difference may be mainly attributed to the difference of the jet kinetic energy, initial Lorentz factor, and medium density among them.
\end{abstract}


\section{Introduction}
It is generally believed that Gamma-ray bursts (GRBs) are originated from the death of massive stars or mergers of compact binaries (e.g., Colgate 1974, Paczynski 1986; Eichler et al. 1989; Narayan et al. 1992; Woosley 1993; MacFadyen \& Woosley 1999; Zhang et al. 2003b; Woosley \& Bloom 2006; Kumar \& Zhang 2015). Prompt gamma-ray lightcurves are highly variable with a duration from milliseconds to thousands of seconds, and their X-ray, optical and radio afterglow usually fade down as a simple power-law up to days, months, and even years (e.g., Kumar \& Zhang 2015). With promptly slewing and precisely locating capacity, the X-ray telescope (XRT) on board the {\em Swift} mission (Gehrels et al., 2004) has observed the very early X-ray emission of a large sample of GRBs triggered with the {\em Swift}/Burst Alert Telescope (BAT). A large fraction of XRT lightcurves show a canonical behavior, as predicted by the external shock models, plus erratic flares from late internal shock emission and an initial steep decaying tail from the last prompt gamma-ray pulse being due to the curvature effect (Zhang et al. 2006; Nousek et al. 2006). A small fraction of XRT lightcurves decay as a single power-law from early to late phases (Liang et al. 2009). In the optical band, about one-third optical lightcurves start with a shallow decay segment and another one-third lightcurves start with a smooth onset bump (e.g., Li et al. 2012; Liang et al. 2013; Wang et al. 2013; Zaninoni et al. 2013; Melandri et al. 2014). Since the early optical afterglow may be less contaminated by prompt optical flares and tails of prompt emission, they could play an unique role in studying the fireball properties and its circum medium (e.g., Liang et al. 2010, 2013). Although the radiation physics of the prompt gamma-rays is still under highly debating, the afterglow fireball models are widely accepted (e.g., Zhang 2014). By considering various effects, such as the energy injection, jet break, medium properties, etc. Wang et al. (2015) suggested that the external shock models can explain the current X-ray and optical afterglow data.

In the framework of the external shock models, the very early afterglow are radiated from reverse shocks (RS) and forward shocks (FS) when the fireball propagates into the circum medium (e.g., M\'{e}sz\'{a}ros \& Rees 1997; Sari et al. 1998; Sari \& Piran 1999a; Zhang \& Kobayashi 2005; Gao et al. 2013). Since bright RS emission was firstly detected in GRB\,990123 (Akerlof et al. 1999), extensive studies on reverse shock emission in the optical/IR bands have been made with the early optical afterglow data (e.g., Sari \& Piran 1999b; Meszaros \& Rees 1999; Fan et al. 2002; Kobayashi \& Zhang 2003b; Zhang et al. 2003a; Kumar \& Panaitescu 2003; Wu et al. 2003; Fan et al. 2004; Nakar \& Piran 2004; Zhang \& Kobayashi 2005; Zou et al. 2005; Jin \& Fan 2007; Harrison \& Kobayashi 2013; Yi et al. 2013; Japelj et al. 2014; Gao et al. 2015; Zhang et al. 2015). Motivated by the extremely bright RS emission detected in GRB\,990123, it is expected that the {\em Swift} optical-UV telescope (UVOT) and ground-based rapid follow-up optical telescopes can detect the RS emission for a large sample of GRBs with XRT prompt and precise localization capacity (e.g. Zhang et al. 2003a). Surprisingly, the detection rate is extremely low (Roming et al. 2006; Li et al. 2012; Japelj et al. 2014; Gao et al. 2015), and plausible RS emission is only occasionally detected in a few GRBs (Melandri et al. 2008; Gomboc et al. 2009; Oates et al. 2009). It was suggested that non-detection of the reverse shock emission may be due to strong suppression of the RS emission in a magnetized outflow (e.g., Zhang\& Kobayashi 2005) or the RS emission peaking at lower frequencies than the optical band (e.g. IR/mm; Mundell et al. 2007b; Melandri et al. 2010; Resmi \& Zhang 2016). It is also possible that the RS emission is overlapped with prompt gamma-rays and it is difficult to be identified (Kopac et al. 2013). Japelj et al. (2014) tried to search for signature of RS emission with a sample of 118 GRBs and identified 10 GRBs with reverse shock signatures -- GRBs\,990123, 021004, 021211, 060908, 061126, 080319B, 081007, 090102, 090424, and 130427A. By modeling their optical afterglow with reverse and forward shock analytic light curves, they found that the physical properties cover a wide parameter range and GRBs with an identifiable reverse shock component show high magnetization parameter $R_B =\epsilon_{\rm B,r}/\epsilon_{\rm B,f}=2-10^4$, where $\epsilon_{\rm B,r}$ and $\epsilon_{\rm B,f}$ are the fractions of internal energy in the RS and FS, respectively. By morphologically analyzing the early optical afterglow lightcurves of 63 GRBs, Gao et al. (2015) found 15 cases with early optical lightcurves dominated by RS emission and derived $R_B\sim 100$.

This paper reports our observations of very bright prompt optical and RS optical emission of GRB\, 140512A (\S 2). We analyze its multi-wavelength data in \S 3 and \S 4. We show that its early optical lightcurve could be attributed to both RS and FS emission from external shock, and we fit the lightcurves with the external shock models by considering both RS and FS components in \S 5. We make comparison of the property of the RS radiation region of GRB\,140512A with that of GRBs\,990123, 090102, and 130427A in \S 6. Discussion and Conclusions are presented in \S 7. Temporal and spectral slopes are defined as $F\propto t^{\alpha} \nu^{\beta}$ and notation $Q_n=Q/10^{n}$ in cgs units is adopted .

\section{Observations and Data Reduction}
GRB\,140512A triggered {\em Swift}/BAT at 19:31:49 UT on 2014 May 12 ($T_0$; Pagani et al., 2014). It also triggered {\em Fermi} Gamma-Ray Burst Monitor (GBM; Meegan et al.2009) at 19:31:42.50 UT (Stanbro 2014) and Konus-Wind at 19:31:50.769 UT (Golenetskii et al.,2014). {\em Swift}/XRT promptly slewed and observed the second gamma-ray peaks since $T_0+98.4$ seconds. Spectroscopic observation with NOT reveals absorption features consistent with FeII and MgII at a common redshift of z=0.725 (de Ugarte Postigo et al. 2014b).

Our optical follow-up with 0.8-m TNT\footnote{TNT is a 0.8-m Tsinghua University - National Astronomical Observatory of China Telescope at Xinglong Observatory runs by a custom-designed automation system for GRB follow-up observations (Zheng et al. 2008). A PI $1300\times1340$ CCD and filters in the standard Johnson Bessel system are equipped for TNT.} in white and $R$ bands started at $T_0+126$ seconds after the BAT trigger (Xin et al. 2014). Our first optical detection is during the second prompt gamma-ray peak. Our data reduction was carried out following the standard routine in
IRAF\footnote{IRAF is distributed by NOAO, which is operated by AURA, Inc., under
cooperative agreement with NSF.} package. Details of our data reduction please refer to Xin et al. (2011). Our observation log and data are reported in Table 1.

We download the BAT data from the NASA {\em Swift} web site. A Python source package  $gtBurst$\footnote{https://github.com/giacomov/gtburst} is used to extract light curves and spectra from the data. The {\em Swift}/XRT lightcurve and spectrum are taken from the
Swift Burst Analyzer (Evans et al. 2010)\footnote{http://www.swift.ac.uk/burst\_analyser/00599037/}. We also download the {\em Fermi}/GBM data of GRB\,140512A from the {\em Fermi} Archive FTP website\footnote{ftp://legacy.gsfc.nasa.gov/fermi/data/}. We extract the lightcurve and spectrum from {\em Fermi}/GBM data with our Python code. Spectral Fitting Package {\em Xspec} is used for our spectral analysis.

\section{Data Analysis}
\subsection{Temporal Analysis}
\begin{figure}[htbp]
 \centering
\includegraphics[angle=0,width=0.8\textwidth]{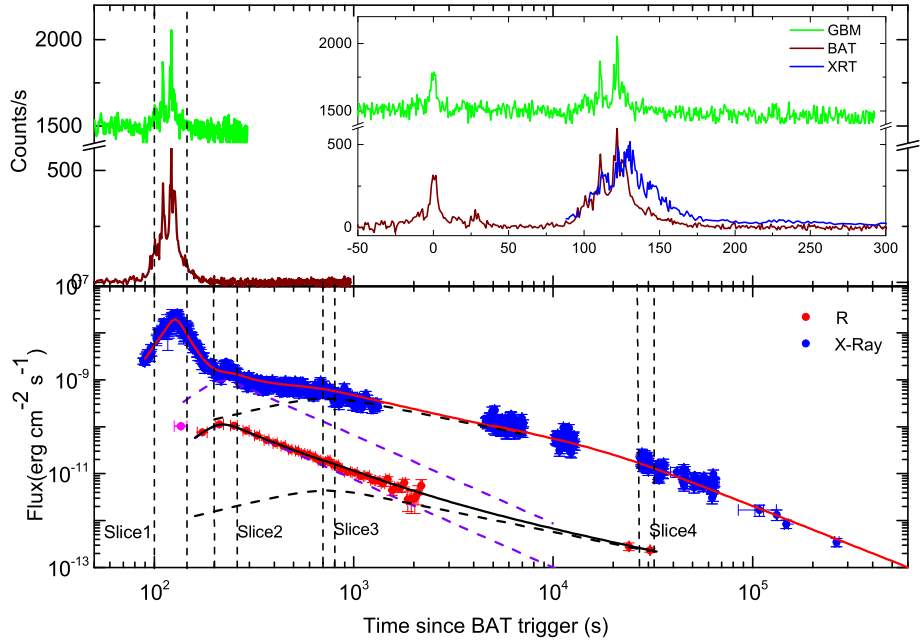}
\caption{Multi-wavelength lightcurves of the prompt and afterglow emission of GRB\,140512A since the second prompt emission episode in the logarithm time scale. The inset of the upper panel shows that prompt X-ray and gamma-ray lightcurves in the linear time scale for illustrating all episodes. Our Empirical fits with a model composing of multiple broken power-law functions for the X-ray and optical data are shown. Each broken power-law component is shown with dashed lines and the sum of these components is shown with solid line. The vertical dashed lines make the time slices of interest for our spectral analysis.}
\label{LCs}
\end{figure}

Figure \ref{LCs} shows prompt and afterglow lightcurves of GRB\,140512A. One can find that the prompt gamma-ray lightcurves observed with {\em Swift}/BAT and {\em Fermi}/GBM are consistent. They have two emission episodes. The first one observed with BAT starts at $T_0-15$ sec, and reaches maximum at $T_0$, then returns to background by $T_0+40$ sec. The second episodes begins at $T_0+85$ sec, reaches maximum at $T_0+120$ seconds, and falls to background by $T_0+170$ seconds. The total duration in the BAT band is $T_{90}=154.8\pm 4.8$ seconds (Sakamoto et al. 2014). The profile of the GBM lightcurve is similar to the BAT lightcurve, but the duration in the GBM band is about 148.0 seconds (Stanbro 2014), which is slightly shorter than that in the BAT band, indicating that the measured $T_{90}$ depends on the energy band and sensitivity of the detectors (Qin et al. 2013).

A bright X-ray flare was detected during the second episode of the prompt gamma-rays. Its profile is also analogue to the BAT lightcurve of the second episode, but has a longer duration than that observed in the BAT and GBM band, as shown in the inset panel of Figure \ref{LCs}. Without considering the fluctuation of the flare, we fit the flare with a smooth broken power-law, which is read
\begin{equation} F=F_0\left [
\left (   \frac{t}{t_p}\right)^{\omega\alpha_1}+\left (
\frac{t}{t_p}\right)^{\omega\alpha_2}\right]^{1/\omega},
\end{equation}
where $\omega$ measures the sharpness of the peak. We get $\alpha_{1,X}=6.27\pm 0.18$, $\alpha_{2,X}=-7.72\pm 0.19$ and $t_{b}=128$ s by fixing $\omega=3$. The rapid increase and decrease of the flux indicate that it would be X-ray emission of the second prompt gamma-ray pulse (see further spectral analysis below). Following the bright X-ray flare, a weak flare was detected at around $T_0+233$ seconds, and then the X-ray lightcurve features as canonical one (Zhang et al. 2006).

A well sampled early optical lightcurve with temporal coverage from the second prompt gamma-ray peak to $T_0+2182$ was obtained from our optical observations. The late optical afterglow were also detected with the 2.5 m Nordic Optical Telescope (NOT) and the GROND telescope and its brightness is $R\sim 19.5$ at $T_0+23940$ seconds (de Ugarte Postigo et al. 2014a) and $r^{'}=19.7\pm 0.1$  at $T_0+30611$ seconds (Graham et al. 2014). For having a broad temporal coverage, these optical data are also included in our analysis, as shown in Figure \ref{LCs}. Noting that the observed optical data are corrected for the Galactic extinction with $A_R=0.348$ and $A_r=0.367$ (Schlafly \& Finkbeiner 2011). Our first optical detection happened during the the second gamma-ray pulse and it may be the result of the prompt optical emission (see spectral analysis below). The optical transient faded down after the first detection and brightened again. The lightcurve continuously faded down after the peak time, featuring as a saddle shaped curve but not a power-law function. We suspect that the early optical afterglow lightcurve is shaped by both the RS and FS emission, similar to that observed in GRB\,990123(Alkerlof et al.1999), GRB\,090102(Steele et al.2009), and GRB\,130427A (Laskar et al. 2013), in which bright RS emission was clearly detected. The bright optical flash would be dominated by the RS emission and the saddle shaped feature around $T_0+10^3$ seconds would be attributed to the emergence of the FS emission. In the X-ray band, a weak bump was also detected at a time around the optical peak. It shows as a shallow decay phase followed by a normal decay phase, then transits to a jet-like decay phase with a slope of $-1.68\pm 0.06$. With the closure relations derived from the standard fireball model (e.g., Zhang et al. 2006), the slope and spectral index of the RS component of GRB\,140512A suggest that its afterglow emission is in the spectral regime between the characteristic frequencies ($\nu_m$ and $\nu_c$) of the synchrotron radiation. in this spectral regime, the decay slope of the pre-break segment is $\alpha=2\beta/2$ for the ISM case. After the jet break and assuming maximized sideways expansion of jets, the lightcurve evolves as $\alpha_j=2\beta-1$. Using the spectral index around the jet break (Slice 4) of -0.57, we infer $\alpha_j=-2.14$, which is steeper than our empirical fit ($\alpha_j=-1.68\pm 0.06$). If the jet sideways expansion
effect can be negligible, $\alpha_j$ is shallower, i.e. $\alpha_j=\alpha-0.75$ for the ISM case (Panaitescu 2005; Liang et al. 2008), yielding $\alpha_j=-1.6$. This is roughly consistent with $\alpha_j$ value derived from our empirical fit, indicaitng that the jet sideways expansion effect is negligible for GRB\,140512A.

Apparently, the X-ray lightcurve behavior after the peak is not consistent with the optical one. We explore whether or not the optical and X-ray lightcurves can be shaped by the RS and FS emission by empirically fit them with a model of multiple smooth broken power laws. Each broken power-law function is taken as Eq. (1). Our strategy is as following.

\begin{itemize}
\item We first fit the optical lightcurve with a model of two smooth broken power-law. Since the first optical data point may be contributed to the prompt optical emission, we do not include it in our temporal analysis. Because only one data point is available before the peak time for the optical afterglow, we fix the peak time at $233$ s. In addition, being due to lack of optical data around $10^{4}$ seconds, we fix the slope around $10^{4}$ seconds as that derived from the X-ray data, i.e., $-0.85$.
\item Fixing the slopes and the peak times as that derived from the fit to the optical data, we fit the X-ray lightcurve with the model by setting the amplitude terms as free parameters in the same time interval as the optical data.

\item To derive the slope of the jet-like decay segment, we fit the XRT lightcurve in the time interval $t>3\times 10^3$ seconds with Eq. 1 by fixing the slope of the normal decaying segment as $-0.85$.
\end{itemize}
Our result fitting curves are shown in Figure \ref{LCs}. Interestingly, the X-ray afterglow lightcurve can roughly fit with our strategy by fixing the slopes as that derived from the optical data in the same time interval. The reduced $\chi^2$ of our fit is 1.41. The early shallow decay segment is shaped by the two emission components. A jet break at $(1.84\pm 0.19)\times 10^4$ is also derived from the X-ray data. These results likely imply that the physical origin of both the X-ray and optical emission are the same.

The rising and decaying slopes are critical to examine the physical origin in the external shock models. In the framework of the RS models for an ISM scenario, the expected RS emission lightcurve increases as $F\propto t^{5}$ and $F\propto t^{0.5}$ for the thin and thick shell cases, respectively, and decays as $t^{\sim -2}$ after the peak time for the two cases. In a wind medium, the decay slope of the RS emission is steeper (about -3 in standard parameters) (Kobayashi \& Zhang 2003a; Zou et al. 2005). The rising and decaying slopes of the optical peak of GRB\,140512A are $3.04\pm 0.09$ and $-1.93\pm 0.07$, respectively. The decaying slope is well consistent with the model prediction for the ISM case. Gao et al. (2015) present a systematic morphological analysis of the GRB early optical afterglow lightcurves. The decaying slope of the RS emission of GRB\,140512A agree with that of the RS II Type in the thin shell case defined by Gao et al. (2015), i.e., $\alpha^{r}_{\rm O, 2}=-(27p+7)/35=-1.94$, but the rising slope ($3.04\pm 0.09$) is shallower than the expectation of the RS II Type, which is $(6p-3)/2=5.25$,  if $p=2.25$, where $p$ is the index of the synchrotron radiating electron spectrum $N_e\propto \gamma_e^{-p}$. Because we have only one data point prior the peak time and it may be also contaminated by the prompt optical emission, the rising index could in fact be steeper than what observed, and more similar to what expected for the RS II Type.

\section{spectral Analysis}
\begin{figure}[htbp]
\includegraphics[angle=0,width=0.5\textwidth]{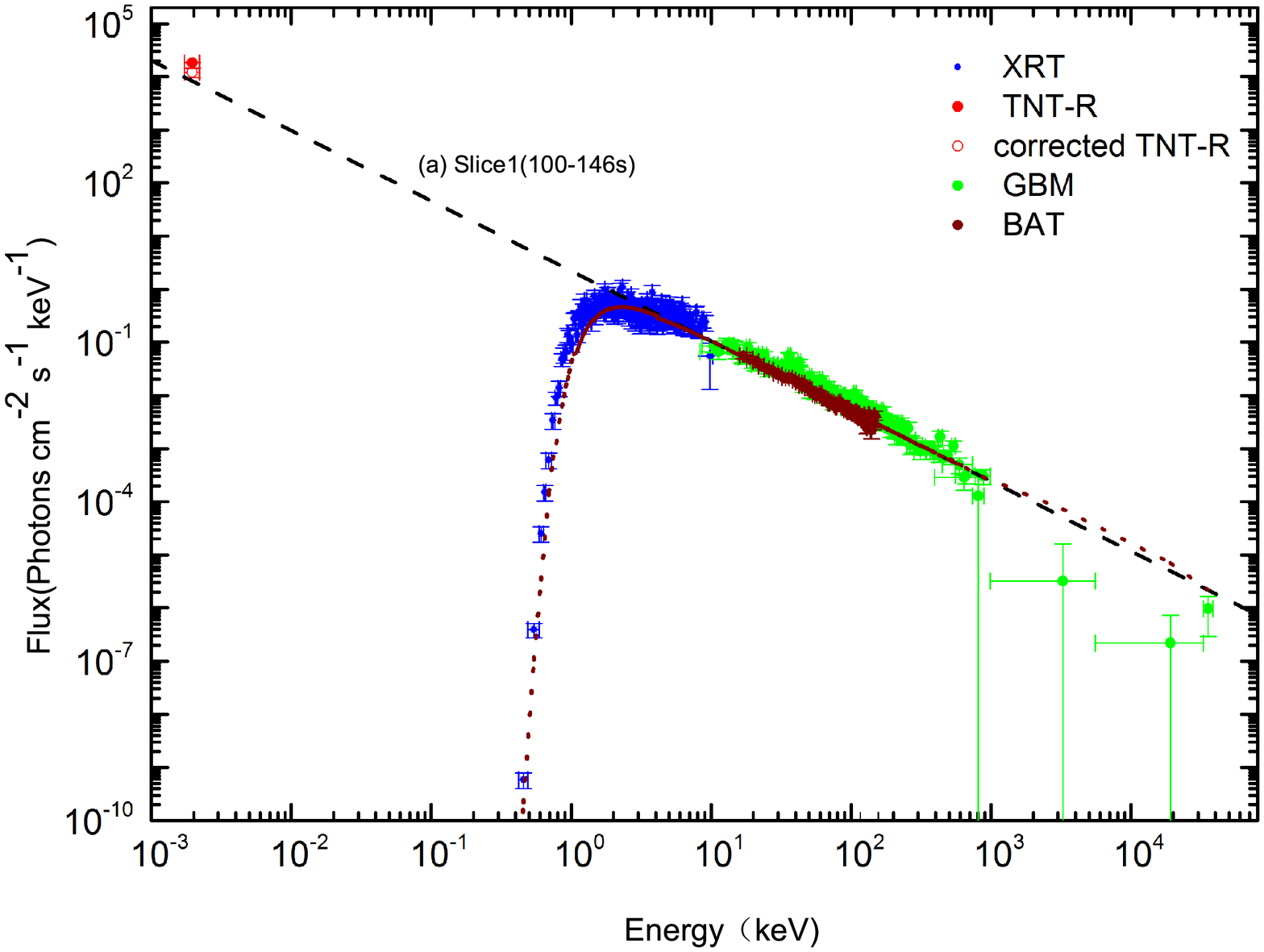}
\includegraphics[angle=0,width=0.5\textwidth]{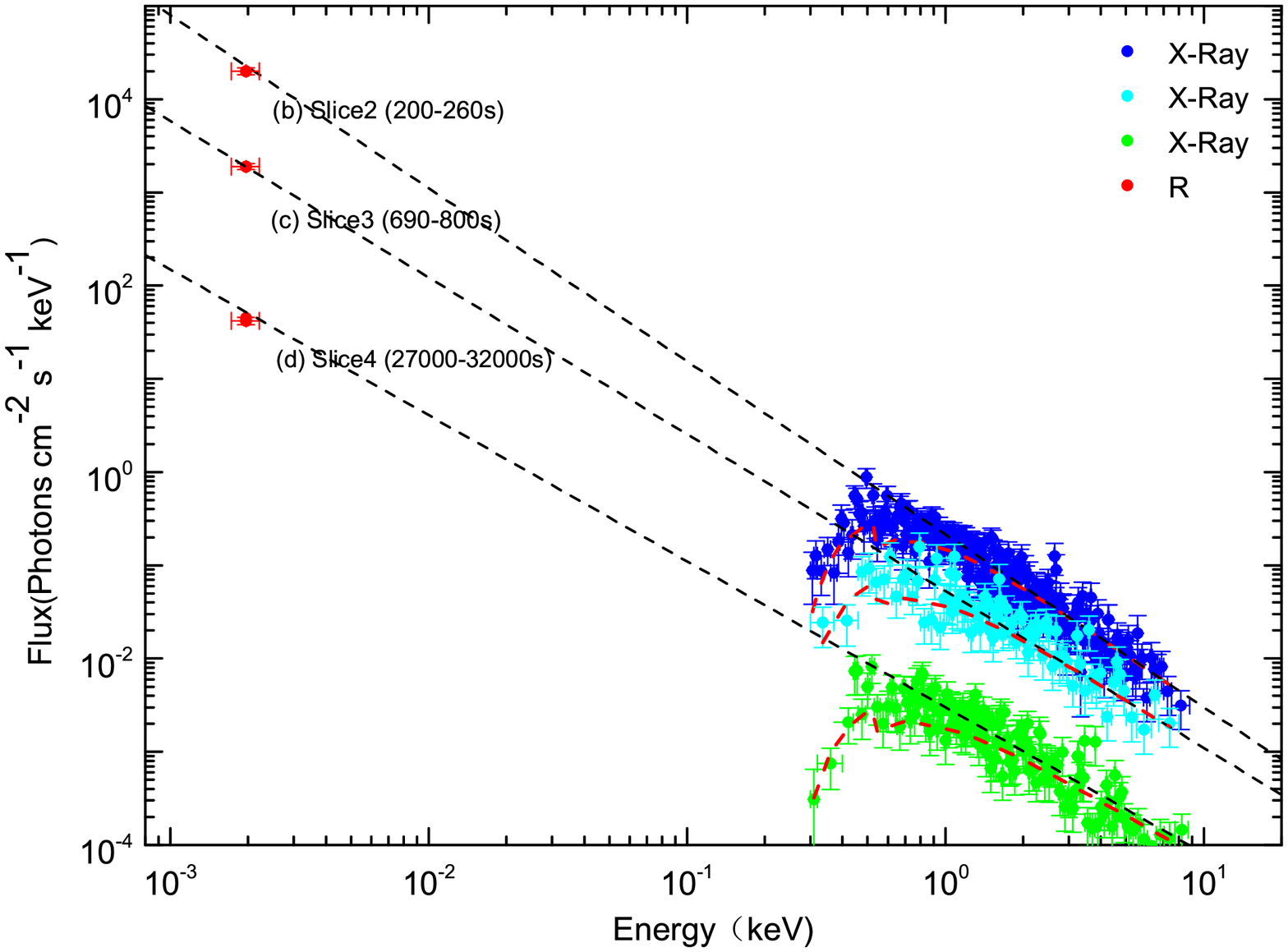}
\caption{Spectral energy distributions of the prompt gamma-ray-Optical-X-ray emission in Slice 1 ({\em left panel}) and the afterglow emissions in Slice 2, 3, and 4 ({\em right panel}). Our joint spectral fits are also shown with dashed lines. For the prompt optical and gamma-ray data, a single power-law model is adequate to fit the broadband spectrum covering from $10^{-3}$ to $10^4$ keV. The prompt optical data marked with a open circle is corrected by removing the contribution of the reverse shock at the same time. An absorbed single power-law function is used for fitting the spectra of the afterglow ({\em dotted line}). \label{SEDs}}
\end{figure}

With simultaneous multi-wavelength observed observations, we present joint spectral analysis for spectra extracted from the data in four time slices as marked in Figure \ref{LCs}. The first time slice is for the prompt optical, X-ray and gamma-ray emission of the second gamma-ray pulse in the time interval [100, 146] s (Slice 1). The other time slices are for the RS peak, the FS peak, and late normal decay segment in the time interval [200, 260] s (Slice 2), [690-800] s (Slice 3), [27000, 32000] s (Slice 4). The joint spectra are shown in Figure \ref{SEDs}. The optical data is corrected for extinction by Milky Way with $E_{B-V}=0.142$ in the burst direction (Schlafly \& Finkbeiner 2011). The neutral hydrogen density of Milky Way in the burst direction is $N_{\rm H}=1.47 \times 10^{21}$ cm$^{-2}$ (Willingale et al. 2013).

Our fits to the spectral energy distributions (SED) of the prompt and afterglow emission in the selected slices are shown in Figure \ref{SEDs} and reported in Table 2. The SED of Slice 1 is derived from the data observed with the TNT optical telescope, XRT, BAT, and GBM, which covers from $10^{-3}$ keV to $3\times 10^4$ keV. The BAT spectrum well agrees with the GBM spectrum in the same energy band coverage. It is interesting that such a broad SED of prompt emission is well fit with a single power-law. The reduced $\chi^2$ is 1.60. The large $\chi^2$ would be due to the calibration of different instruments (GBM, BAT, XRT, and optical telescopes). The derived photon index is $\Gamma_\gamma=-1.32\pm 0.01$\footnote{Stanbro (2014) reported that the time-averaged GBM spectrum can be fit
by a cutoff power law model, which yields a photon index of $-1.33\pm 0.03$ and peak energy of $588\pm 84$ keV. The photon index is well consistent with ours. Inspecting the SED shown in Figure \ref{SEDs}, a plausible break with large uncertainty of the data at around several hundreds of keV is indeed observed.}.  Noting that the optical extinction and neutral hydrogen absorbtion of the GRB host galaxy are taken into account, but they are negligible. The optical flux is slightly higher than our spectral fit line, even by subtracting possible contamination of the rising part of the reverse shock emission with $F\propto t^{3.04\pm 0.09}$ (shown in  Figure \ref{SEDs} with an open circle).

The selection of the SED of Slice 2 is for the peak of the optical afterglow. Noting that a weak X-ray flare like event is also simultaneously detected during the optical peak. We find that an absorbed single power-law function is adequate to fit the joint optical and X-ray spectrum without considering the host galaxy extinction on the R band data. The derived photon index is $-1.86\pm 0.01$, implying that the optical and X-ray peak in this time slice may have the same physical origin, says, the RS emission of the GRB fireball.

The selection of the SED of Slice 3 is for the peak time of the possible FS emission. Our empirical fit suggests the FS emission may peak at around this time interval. The X-ray lightcurve also transits to a steeper decay slope after this time slice. We find the optical and X-ray spectrum also can be well fit with a single power-law, yielding a photon index of $-1.68\pm 0.01$. Based on our empirical analysis shown in Figure \ref{LCs}, we can find that the X-ray emission in this time slice may be dominated by the FS emission, but the optical emission may still dominated by the RS emission.

The selection of the SED of Slice 4 is for late FS emission around the jet break. It is also found that the optical and X-ray emission component can be fitted with an absorbed power-law, with a photon index of $-1.57\pm 0.01$. The spectrum is even harder than that in Slice 2 and Slice 3. The spectral hardening observed in Slices 2-4 would be due to the competition between the RS and FS emission.

\section{Theoretically Modeling the Afterglow Lightcurves}
Our analysis above suggests that the optical and X-ray afterglow may be attributed to the RS and FS emission of external shocks. We present in this section our fits to the X-ray and optical afterglow lightcurves with the standard external shock models. The details of the forward shock model please refer to Sari et al. (1998), Huang et al. (2000), and Fan et al. (2006). The reverse shock model please refer to Yi et al. (2013) and Gao et al. (2015). We assume that the spectra of radiating electrons in both the forward and reverse shock regions are $N_e\propto \gamma_e^{-p}$. With the observed spectral index and temporal decay slope of the normal decay segment, we suggest that both the optical and X-ray emission should be in the spectral regime between $\nu_m$ and $\nu_c$, and take $p=2\beta+1=2.5$, where we roughly take $\beta$ as the average of the spectral indices of the afterglow. The fractions of internal energy to the electrons and magnetic field are $\varepsilon_{\rm e,r}$ and $\varepsilon_{\rm B,r}$ in the reverse shock region and $\varepsilon_{\rm e,f}$, and $\varepsilon_{\rm B,f}$ in the forward shock region. Our empirical analysis shows that the rising and decaying slope of the RS emission is consistent with the expectations in the ISM scenario. We then take an constant medium density ($n$). The temporal evolution of both minimum and cooling frequencies ($\nu_m$ and $\nu_c$) in the reverse and forward shock regions are taken from Rossi et al. (2003), Fan et al. (2006), Zhang et al. (2007) and Yi et al. (2013).

\begin{figure}[htbp]
 \centering
\includegraphics[angle=0,width=0.4\textwidth]{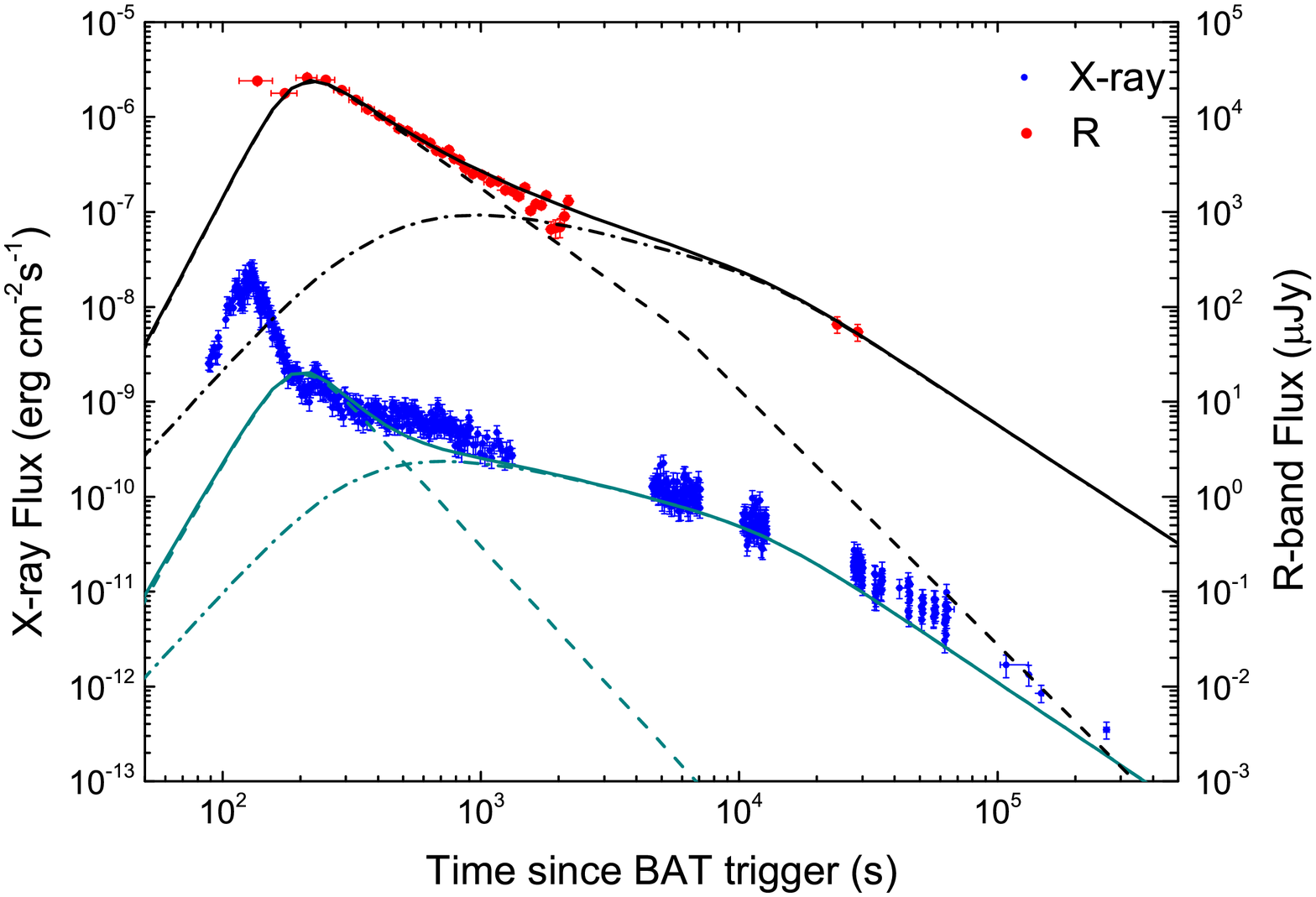}
\includegraphics[angle=0,width=0.4\textwidth]{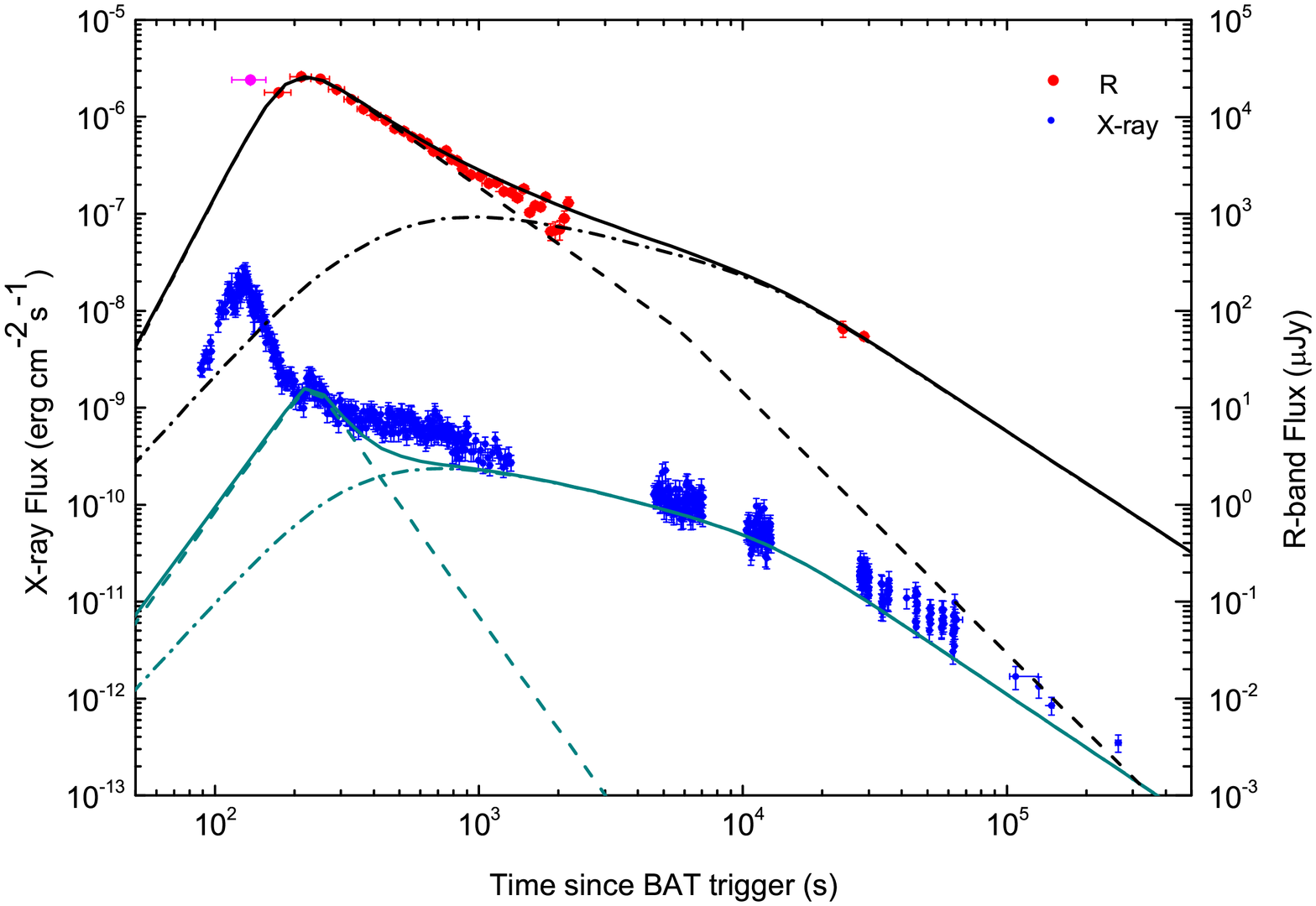}
\caption{{\em Left panel---}Theoretically fits (solid lines) to the optical and X-ray afterglow lightcurves with external shock models by considering emission from both reverse (dashed lines) and forward shocks (dot-dashed lines).{\em Right panel---}the same as the left penal, but the weak X-ray peak around $T_0+200$ seconds is interpreted as an X-ray flares superimposed to the afterglow phase. \label{Model_Fit}}
\end{figure}

\begin{figure}[htbp]
\includegraphics[angle=0,width=0.3\textwidth]{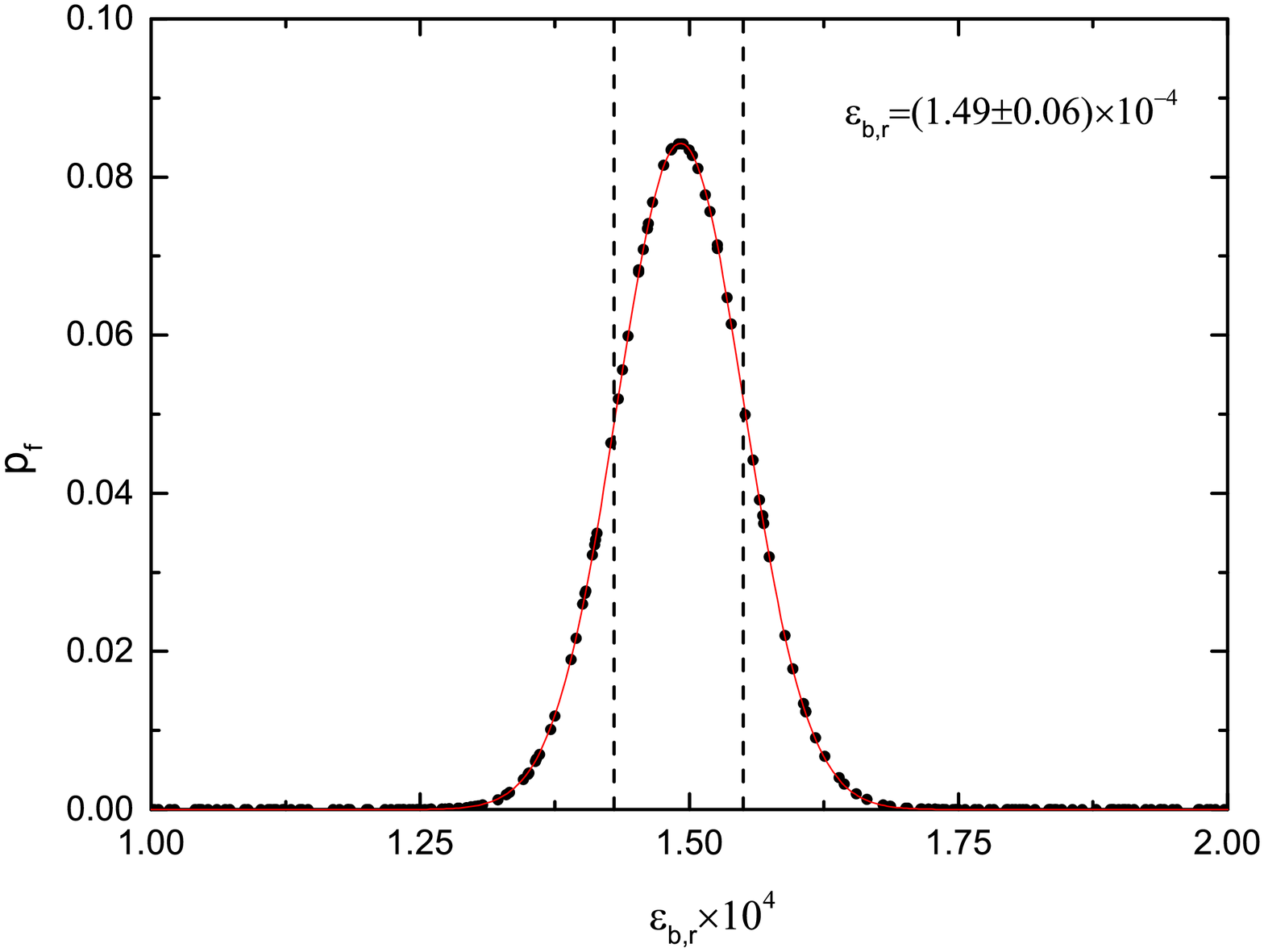}
\includegraphics[angle=0,width=0.3\textwidth]{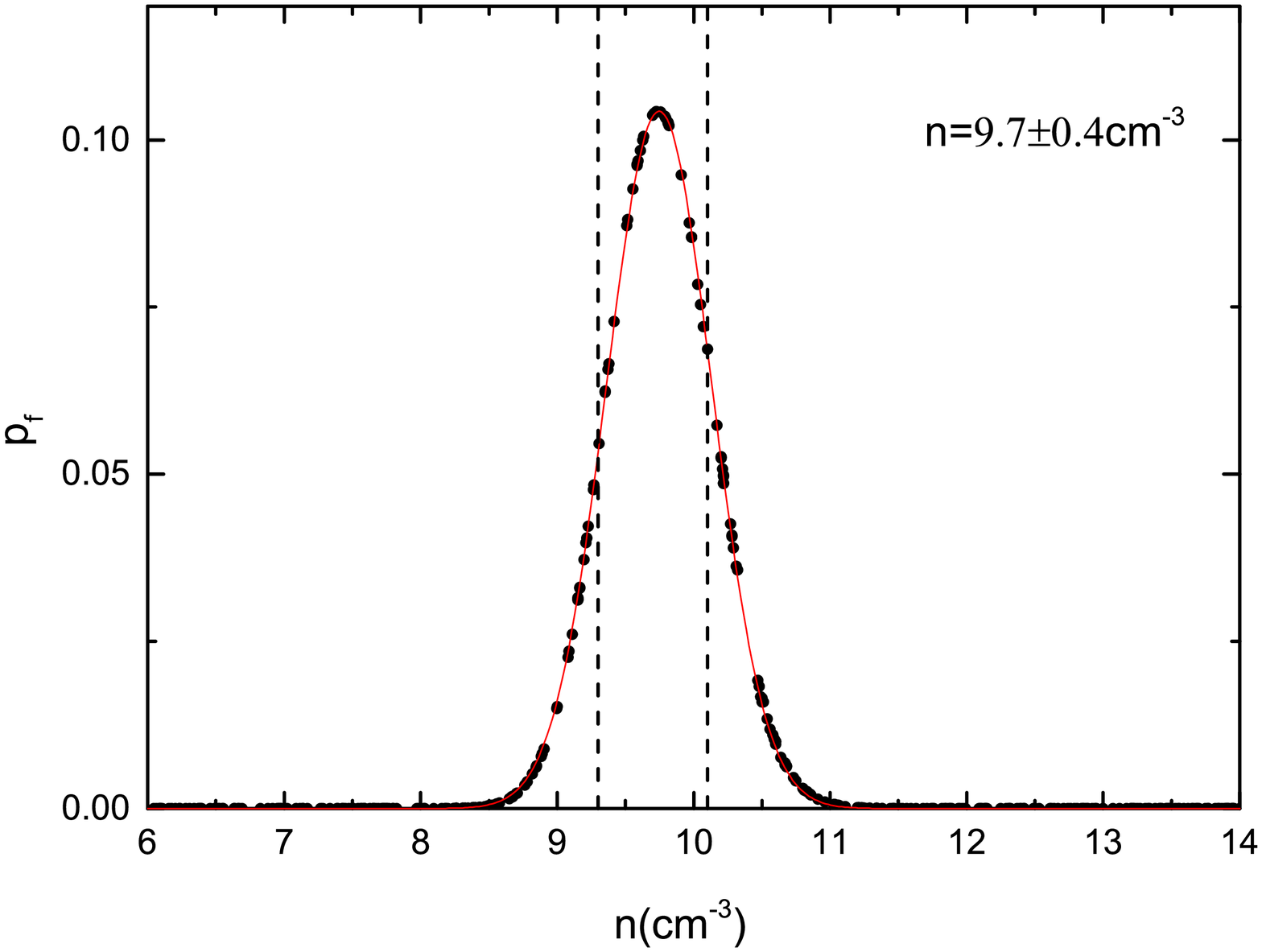}
\includegraphics[angle=0,width=0.3\textwidth]{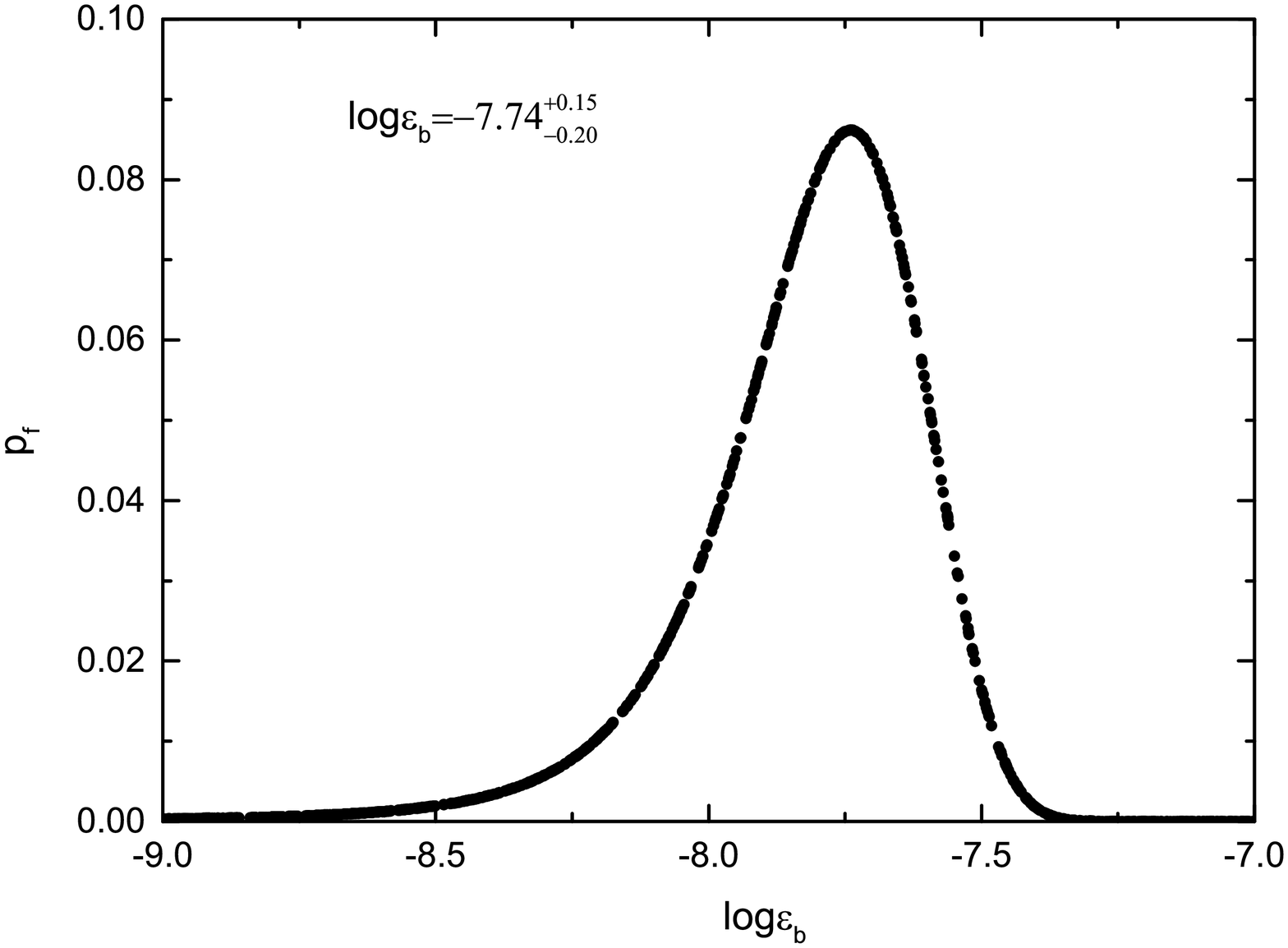}
\includegraphics[angle=0,width=0.3\textwidth]{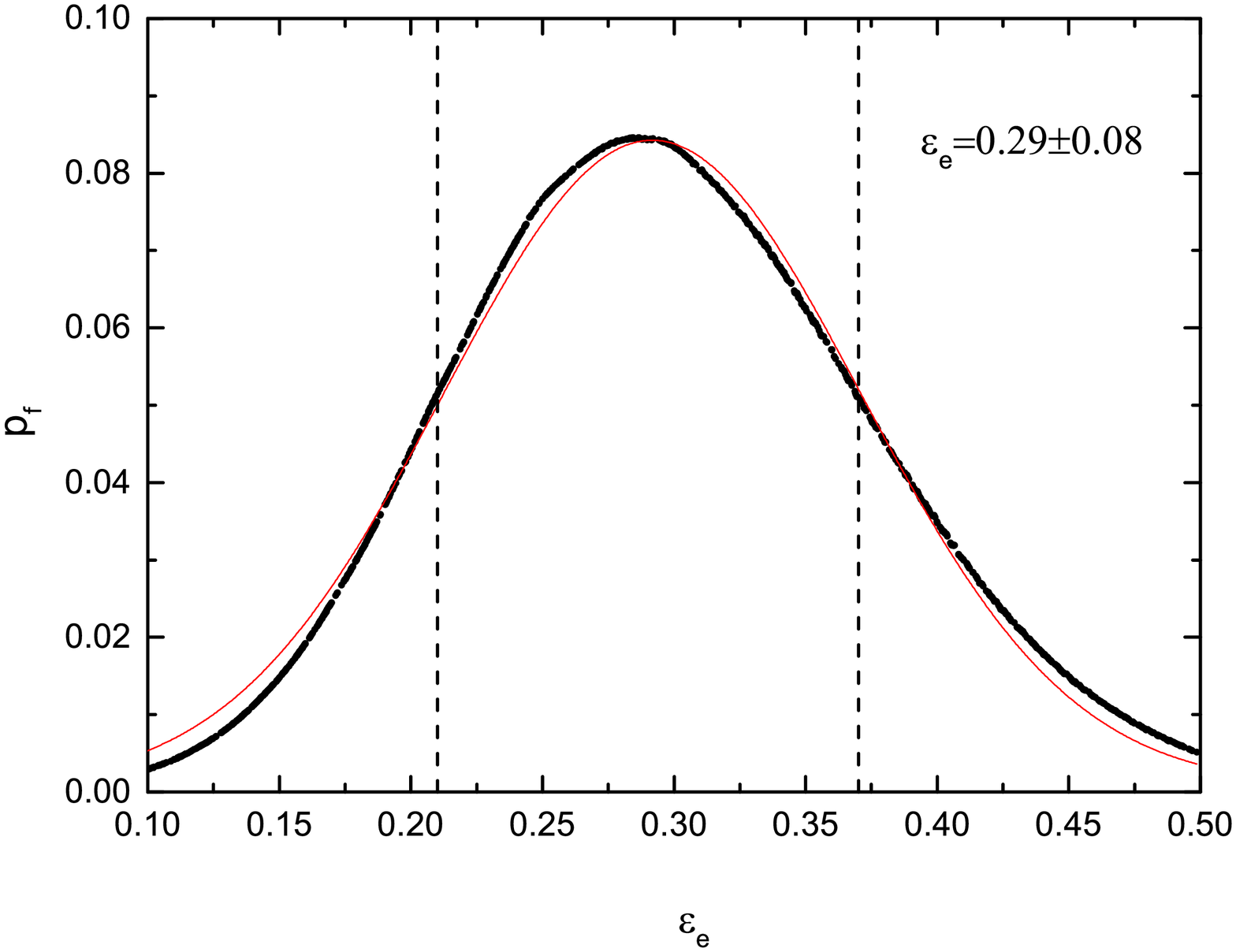}
\includegraphics[angle=0,width=0.3\textwidth]{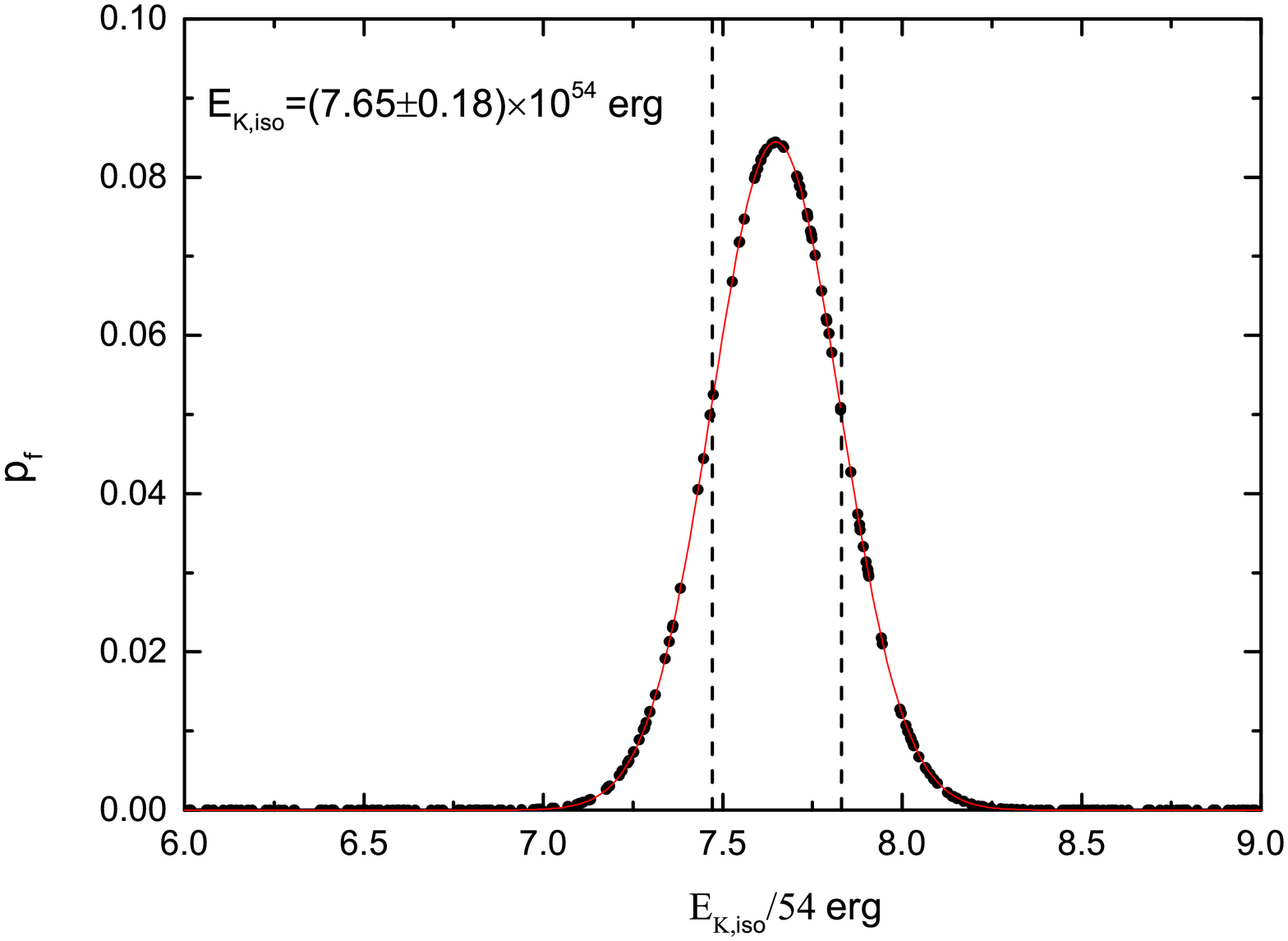}
\includegraphics[angle=0,width=0.3\textwidth]{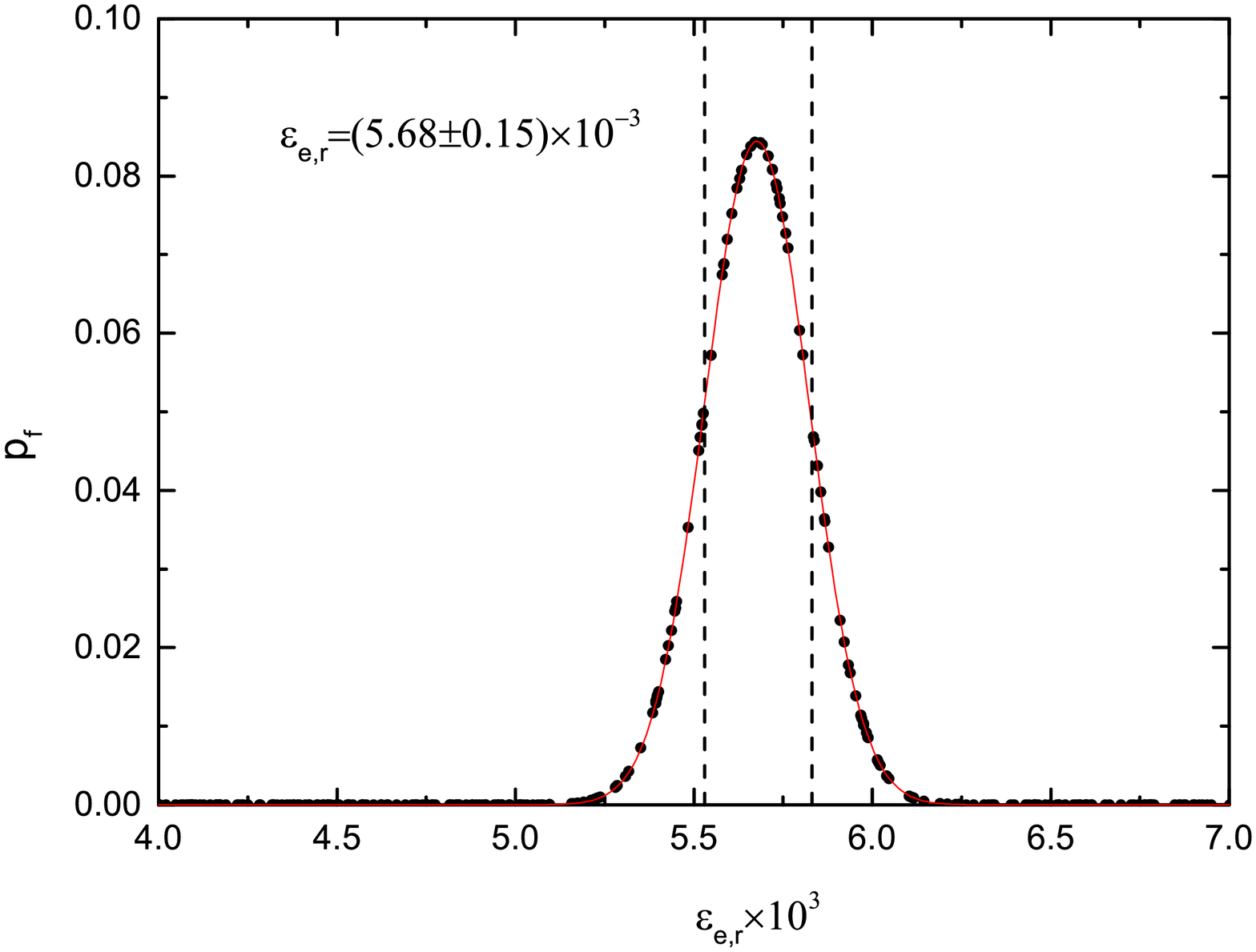}
\includegraphics[angle=0,width=0.3\textwidth]{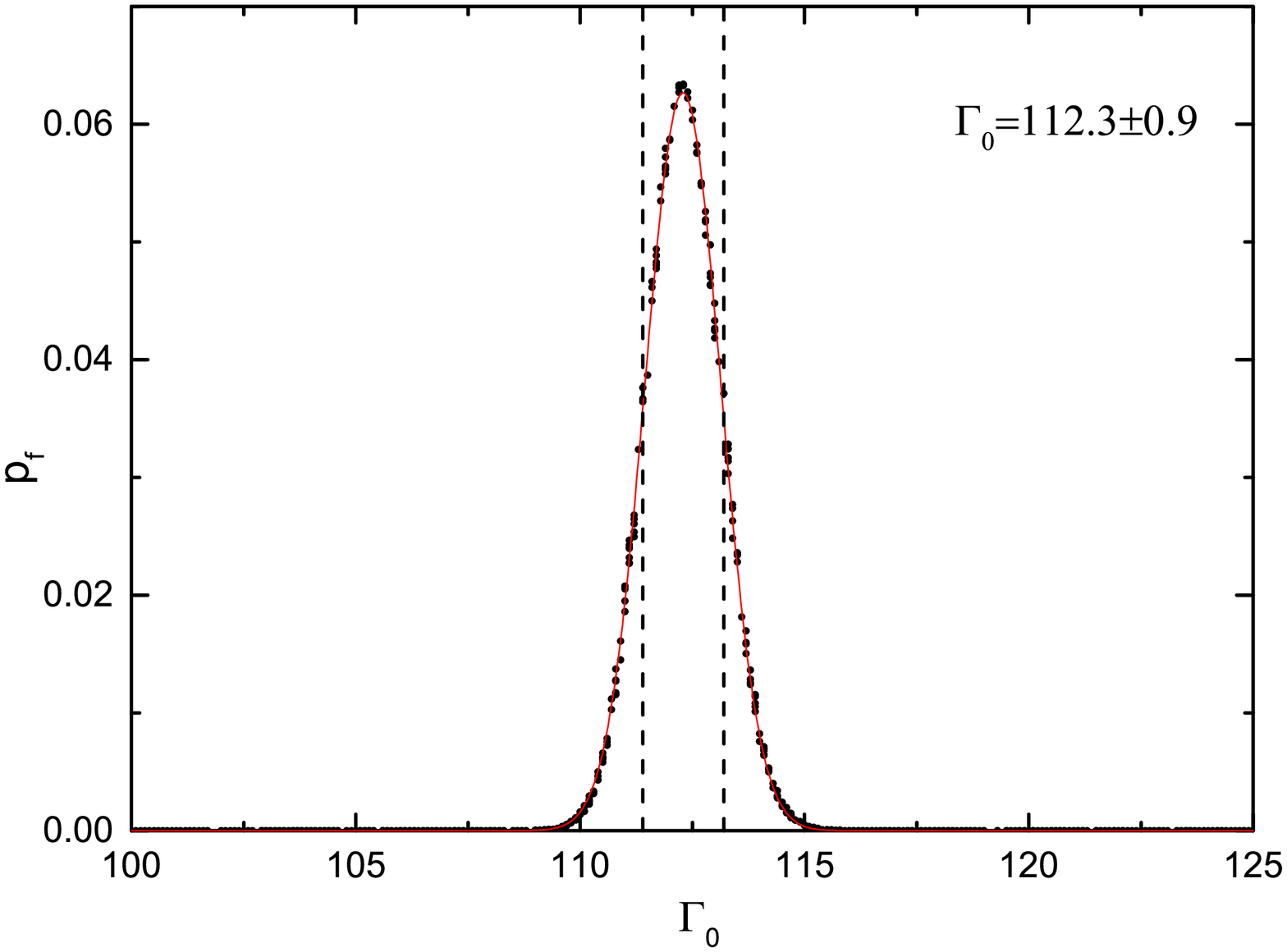}
\includegraphics[angle=0,width=0.3\textwidth]{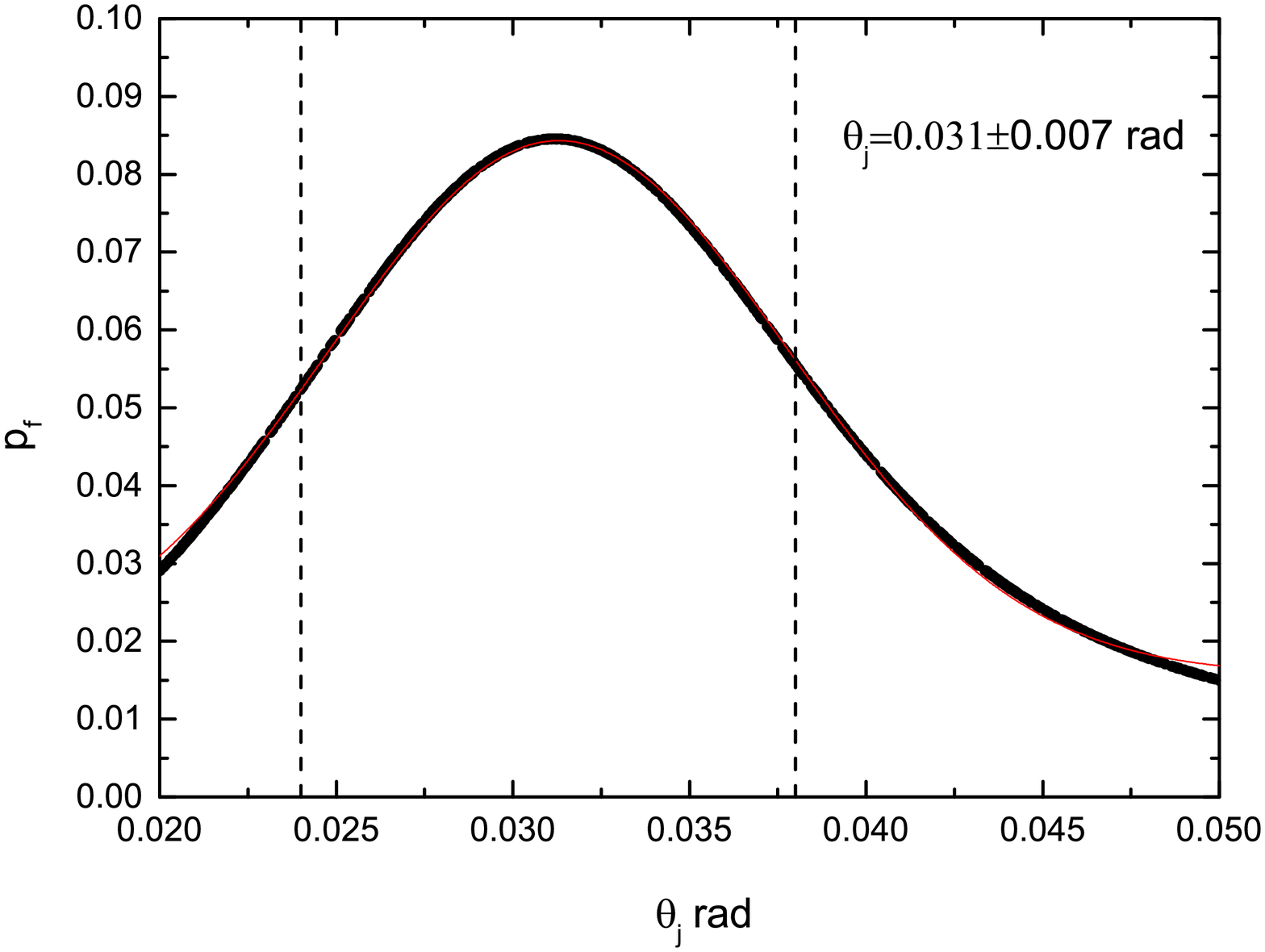}
\caption{Probability distributions of the afterglow model parameters along with our Gaussian function fits (solid red lines) for GRB\,140512A. The dashed vertical lines mark the $1\sigma$ confidence level of the parameters in this parameter set. }\label{Paraeters}
\end{figure}

The free model parameters include $\varepsilon_{\rm e,r}$, $\varepsilon_{\rm B,r}$, $\varepsilon_{\rm e,f}$, $\varepsilon_{\rm B,f}$, $n$, $\Gamma_0$ (the initial fireball Lorentz factor), $\theta_j$ (jet opening angle), and $E_{\rm K,iso}$ (the kinetic energy of the fireball). We use an MCMC technique to make the best fit to the observed lightcurves. The details of the technique and our procedure please see Xin et al. (2016). Our results are reported in Table 3 and shown in Figure \ref{Model_Fit}(a). The $1\sigma$ errors of the parameters are shown in Figure \ref{Paraeters}. It is found that the standard external shock models can well fit the lightcurves by considering both the RS and FS emission.

Gao et al. (2015) reported that typical GRBs usually have an $\epsilon_{\rm B, f}$ value being much lower than the range of $[10^{-2}, 10^{-6}]$. We derive $\epsilon_{\rm B, f}=1.82\times 10^{-8}$, which is extremely low. Noting that for a constant density medium, the cooling frequency of synchrotron emission frequency is given by $\nu_c =6.3 \times 10^{15} {\rm\ Hz}
(1+z)^{-1/2} (1+Y)^{-2} \epsilon_{B,-2}^{-3/2}E_{\rm K,iso, 52}^{-1/2} n^{-1}
t_d^{-1/2}$ (Sari et al. 1998; Yost et al. 2003), where $Y$ is the Inverse Compton scattering parameter and $t_d$ is the observer's time in unit of days. One can see that $\nu_c$ is sensitive to $\epsilon_B$. As time increases, $\nu_c$ is getting smaller. The extremely low $\epsilon_B$ ensures that both the optical and X-ray emission is still in the regime $\nu<\nu_c$ at late epoch in the dense medium. With the model parameters of both RS and FS emission from GRB\,990123 (Panaitescu \& Kumar 2001), Fan et al. (2002) proposed that the magnetization parameter of RS and FS regions should be different. Defining magnetization parameter with $R_B\equiv \epsilon_{\rm B, r}/\epsilon_{\rm B, f}$, we get $R_B=8187$ for GRB\,140512A. Noting that the estimated $R_B$ values are dramatically different among bursts (e.g., Japelj et al. 2014; Gao et al. 2015). The $R_B$ value of GRB\,140512A derived in our analysis is at the high end of the range obtained by Japelj et al. (2014). The ratio $R_e\equiv\epsilon_{\rm e, r}/\epsilon_{\rm e, f}$ may indicate the relative radiation efficiency of the reverse shocks to the forward shocks. We get $R_e=0.02$, likely implying that the radiation efficiency of reverse shocks ia much lower than the forward shocks.

\section{Comparison of the RS emission of GRB\,140512A with GRBs\,990123, 090102, and 130427A}
\begin{figure}[htbp]
 \centering
\includegraphics[angle=0,width=0.8\textwidth]{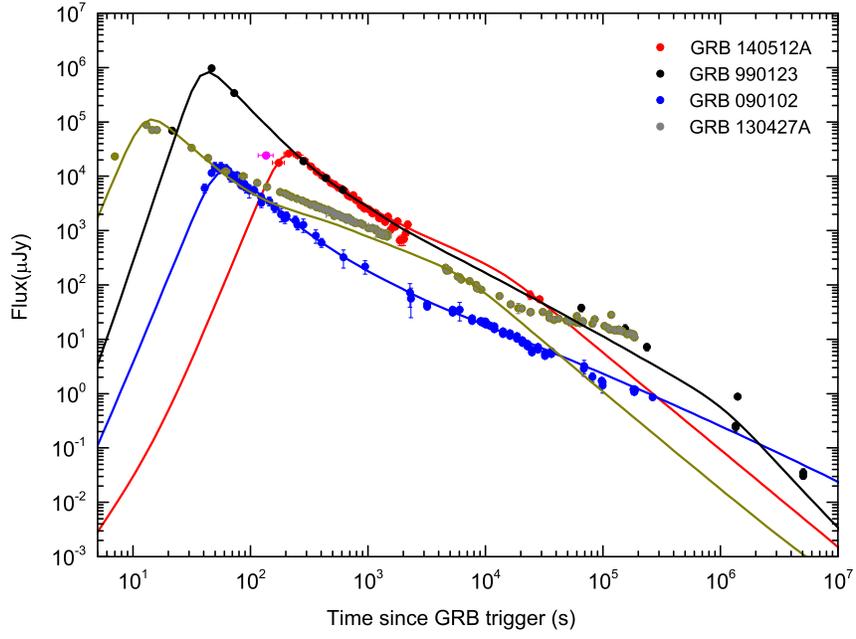}
\caption{Comparison of the afterglow lightcurve GRB\,140512A with GRBs 990123, 090102, and 130427A. Our external model fits by considering both the FS and RS emission are also shown (lines). \label{Comparison}}
\end{figure}

Bright RS emission was detected in the early optical afterglow of GRBs\,090102(Steele et al.2009), 990123(Akerlof et al.1999), and 130427A (Laskar et al. 2013). We compare the optical afterglow lightcurve of GRB\,140512A with these GRBs in Figure \ref{Comparison}. One can observe that their shapes are quite similar at $T_0+t<10^4$ seconds, with different peak time and peak luminosity. This fact suggests that they may be produced in radiative regions with similar micro-physical conditions and the difference would be due to the kinetic energy, Lorentz factor, and surrounding medium. Therefore, we investigate whether the micro-physical parameters ($\epsilon_{\rm e, r}$ and $\epsilon_{\rm B, r}$) of their RS radiation regions are similar. We fit the early optical lightcurves by keeping $\epsilon_{\rm e, r}$ and $\epsilon_{\rm B, r}$ the same as that derived from GRB\,140512A and varying the parameters of $E_{\rm K, iso}$, $n$, $p$, and $\Gamma_0$. The parameters of the FS radiation regions, including micro-physics parameters ($\epsilon_{\rm e, f}$ and $\epsilon_{\rm B, f}$) and jet opening angle, also vary for making fits to the late optical lightcurves. Our fitting curves are also shown in Figure \ref{Comparison}. One can find that the RS emission of these GRBs can be modeled by taking the same $\epsilon_{\rm B, r}$ value as that of GRB 140512A, i.e., $\epsilon_{B, r}=1.49\times 10^{-4}$. The $R_B$ values of GRBs\,990123, 090102, and 130427A are similar to that of GRB\,140512A. The $\epsilon_{e, r}$ values of GRBs\, 090102 and 130427A are also the same as that of GRB 140512A, but the $\epsilon_{e, r}$ value of GRBs\,990123 is much larger than the other GRBs, yielding a much larger value of $R_e$ (=0.4) for GRB\,990123. It is much larger than that of other GRBs. This may suggest the extremely bright reverse shock emission of GRB\,990123 is due to its high radiation efficiency of its reverse shocks. These results may suggest that the apparent difference of the RS emission in these GRBs may be mainly attributed to the difference of the jet kinetic energy, initial Lorentz factor, radiation efficiency, and medium density among them.

\section{Conclusions and Discussion}
We have reported our very early optical observations of GRB\,140512A and analyze its multi-wavelength data by using our data together with data observed with the {\em Swift} and {\em Fermi} missions. We summary our results as following.
\begin{itemize}
\item We obtain very bright and well sampled optical lightcurve with a temporal coverage from 136 sec to about 8 hours after the burst trigger.
 \item The joint optical-X-ray-gamma-ray prompt emission spectrum can be fit by a single power-law with index $-1.32\pm 0.01$, Our fit also shows that the optical extinction and neutral hydrogen absorbtion of the GRB host galaxy are negligible. This may result in detection of very bright optical emission ($R=13.09$ mag at $T_0+136$ seconds during the second pulse of the prompt gamma-rays).
 \item Our empirical fit to the afterglow lightcurves indicates that the observed bright optical afterglow, which reached $R=13.00$ mag at the peak time, is consistent with predictions of the RS and FS emission of external shock models.
  \item Joint optical-X-ray afterglow spectrum is well fitted with an absorbed single power-law, with an index evolving with time from $-1.86\pm 0.01$ at the peak time to $-1.57\pm 0.01$ at late epoch, which could be due to the evolution of the ratio of the RS to FS emission fluxes.
 \item  Fitting the lightcurves with standard external models, we derive the physical properties of the outflows and find $R_B\equiv\epsilon_{\rm B, r}/\epsilon_{\rm B, f}=8187$ indicating a high magnetization degree in the RS region. We also find that $R_e= 0.02$, implying the radiation efficiency of the RS is much lower than that in FS.
 \item The $R_B$ value of GRB\,140512A is similar to that of GRBs\,990123, 090102, and 130427A. Their apparent difference would be mainly due to the difference of their jet kinetic energy, initial Lorentz factor and medium density among them. A large $R_e$ value in GRB\,990123 may also suggest the extremely bright reverse shock emission of GRB\,990123 is due to the high radiation efficiency of its reverse shock.
\end{itemize}

 Our results indicate that the early, bright optical emission of GRB 140512A can be well explained with the RS model. More important, we find that $\epsilon_{\rm B,r}$ value is much larger than $\epsilon_{\rm B,f}$ for GRB\,140512A, similar to that in GRB\,990123 (Fan et al. 2002). To explain bright GRB\,990123-like reverse shock emission, Zhang et al. (2003a) found that the reverse shock should be more magnetized than the forward shock. Noting that a strong reverse shock is developed if the outflow from the central engine is baryonic. The detection of bright RS emission likely suggests that a moderately magnetized reverse shock in which the magnetic field is strong enough to enhance the reverse shock emission but not strong enough to suppress reverse shock dynamics (Zhang \& Kobayashi 2005). This favors the reverse shock detection in a good fraction of GRBs (e.g. Gomboc et al. 2008; Harrison \& Kobayashi 2013; Japelj et al. 2014; Gao et al. 2015). Since the upstream of RS is the ejecta from the central engine, a high $R_B$ value of GRB\,140512A may hint a strongly magnetized central engine of this GRB (e.g., Fan et al. 2002; Zhang et al. 2003a; Zhang \& Yan 2011; L\"{u} et al. 2014; 2015).

In this analysis we explain the weak X-ray flare simultaneously detected around the optical RS emission peak of GRB\,140512A as RS X-ray emission, based on the fact that the joint optical and X-ray spectrum can be well fit with a single power-law. However, we should note that RS emission is expected to be bright in the optical and radio bands since the temperature of the RS region is low (e.g. Resmi \& Zhang 2016). This gives rise an issue on explanation of the X-ray flare as the RS emission. Since prompt X-ray flares are usually detected in the early afterglow phase (e.g., O'Brien et al. 2006; Liang et al. 2006; Chincarini et al. 2007; Peng et al. 2014; Yi et al. 2016), we replace the RS emission in the X-ray band from our model with our empirical fit to the flare. The result is presented in Fig. 3(right panel). One can observe that it roughly represents the observed X-ray light curve. Therefore, one cannot exclude the possibility of internal origin of the weak X-ray flare(e.g., Liang et al. 2006).

\section{Acknowledgement}
This work is supported by the National Basic Research Program of China (973 Program, grant No. 2014CB845800),
the National Natural Science Foundation of China (Grant No. 11533003, 11103036 and U1331101), the Strategic Priority Research
Program ¡°The Emergence of Cosmological Structures¡± of the Chinese Academy of Sciences (grant XDB09000000), the Guangxi Science
 Foundation (Grant No. 2013GXNSFFA019001). We very appreciate helpful comments/suggestions from the referee. We also acknowledge the use of the public data from the Swift data archive.
We appreciate helpful discussion with Bing Zhang, He Gao, Hou-Jun L"{U}, and Xue-Feng Wu.

\newpage
\begin{deluxetable}{cccccc}
\tabletypesize{\small}
\tablewidth{0pt}
\label{Tab:pub-data}
\tablecaption{Optical Afterglow Photometry Log of GRB\,140512A}
\tablehead{
\colhead{$T-T_0$(mid,sec)} &
\colhead{Exposure (sec)} &
\colhead{Mag} &
\colhead{Merr} &
\colhead{Filter} &
\colhead{Telescope}
}
\startdata
136   &  20  &  13.090   &  0.009  &  W  &  TNT  \\
174   &  20  &  13.417   &  0.011  &  W  &  TNT  \\
212   &  20  &  13.000   &  0.009  &  W  &  TNT  \\
251   &  20  &  13.068   &  0.009  &  W  &  TNT  \\
289   &  20  &  13.335  &  0.010  &  W  &  TNT  \\
328   &  20  &  13.590  &  0.013  &  W  &  TNT  \\
366   &  20  &  13.833   &  0.016  &  W  &  TNT  \\
404   &  20  &  14.000  &  0.017  &  W  &  TNT  \\
443   &  20  &  14.129  &  0.020  &  W  &  TNT  \\
481   &  20  &  14.340  &  0.024  &  W  &  TNT  \\
520   &  20  &  14.412  &  0.027  &  W  &  TNT  \\
558   &  20  &  14.556  &  0.032  &  W  &  TNT  \\
596   &  20  &  14.615   &  0.031  &  W  &  TNT  \\
635   &  20  &  14.728   &  0.038  &  W  &  TNT  \\
673   &  20  &  14.920  &  0.046  &  W  &  TNT  \\
711   &  20  &  14.982  &  0.049  &  W  &  TNT  \\
750   &  20  &  14.916   &  0.044  &  W  &  TNT  \\
788   &  20  &  15.133  &  0.053  &  W  &  TNT  \\
827   &  20  &  15.166  &  0.062  &  W  &  TNT  \\
865   &  20  &  15.376   &  0.076  &  W  &  TNT  \\
932   &  60  &  15.533   &  0.071  &  R  &  TNT  \\
1010  &  60  &  15.563   &  0.078  &  R  &  TNT  \\
1088  &  60  &  15.754  &  0.099  &  R  &  TNT  \\
1166  &  60  &  15.735   &  0.094  &  R  &  TNT  \\
1244  &  60  &  15.972   &  0.111  &  R  &  TNT  \\
1322  &  60  &  15.992  &  0.125  &  R  &  TNT  \\
1401  &  60  &  16.135  &  0.143  &  R  &  TNT  \\
1479  &  60  &  15.899  &  0.115  &  R  &  TNT  \\
1557  &  60  &  16.504   &  0.232  &  R  &  TNT  \\
1635  &  60  &  16.329  &  0.204  &  R  &  TNT  \\
1713  &  60  &  16.364  &  0.230  &  R  &  TNT  \\
1791  &  60  &  16.106   &  0.208  &  R  &  TNT  \\
1869  &  60  &  16.994  &  0.489  &  R  &  TNT  \\
1947  &  60  &  16.965   &  0.559  &  R  &  TNT  \\
2026  &  60  &  16.937  &  0.570  &  R  &  TNT  \\
2104  &  60  &  16.655  &  0.473  &  R  &  TNT  \\
2182  &  60  &  16.262   &  0.386  &  R  &  TNT  \\
23940 & 300  &  19.5     &    -    &   R &  NOT  \\
30611 & 264  &  19.7   &  0.1   &   r'    &  GROND\\
\enddata
\tablecomments{
The reference time $T_0$ is {\em Swift} BAT trigger time.
``$T-T_0$" is the middle time of the observations. ``Exposure" is the exposure time in second.
``Merr" means the uncertainty of magnitude. $W$ and $R$ band data are calibrated by nearby USNO B1.0 R2 magnitude.
All Data are not corrected for the Galactic extinction, which is $E_{B-V}=0.142$ at the burst direction (Schlafly \& Finkbeiner 2011).}
\end{deluxetable}

\clearpage
\begin{table}
\footnotesize
\tablewidth{450pt}
\caption{Results of our Empirical fits to the X-ray and and optical afterglow lightcurve with a multiple smooth broken power-law model utilizing a strategy described in \S 3.1.  }
\centering
\begin{tabular}{lllllll}
\hline\hline
  Band  & $F^{r}_{\rm 0}(\rm erg\ cm^{2}\ s^{-1})$ & $\alpha^{r}_{1}$ &  $\alpha^{r}_{2}$  &  $t^{r}_{p}(s)$    &$-$&   $-$ \\
\hline
  Optical  & $(1.41\pm0.02)\times10^{-10}$  &   $3.04\pm0.09$  &   $-1.93\pm0.07$  &  -    &$-$&  $-$   \\
  X-ray   & $(1.31\pm0.05)\times10^{-9}$  &   $3.04$(fxied)   &   $-1.93$ (fixed)   &  $233$    &$-$&  $-$  \\
\hline
  Band  &  $F^{f}_{\rm 0}(\rm erg\ cm^{2}\ s^{-1})$  &  $\alpha^{f}_{1}$  &  $\alpha^{f}_{2}$  &  $t^{f}_{p}(s)$   &   $\alpha_{\rm j}$   &   $t^{f}_{j}$   \\
\hline
  Optical &  $(5.90\pm2.04)\times10^{-12}$    &   $0.85\pm 0.06$  &   $-0.85$ (fixed)  &  $606\pm138$  &  $-$   & $-$  \\
  X-ray &  $(5. 65\pm0.12)\times10^{-10}$  &   $0.85$ (fixed) &   $-0.85$ (fixed)  & $606$ (fixed)  & $-1.68\pm0.06$ &$(1.84\pm0.19)\times10^{4}$ \\
\hline
\end{tabular}
\tablecomments{The superscripts ``r" and ``f" stand for the possible RS emission and FS emission parts, respectively, and subscript ``j" is for the post jet break segment. The reduced $\chi^2$ of our fits are 1.61 and 1.41 for the to the optical and X-ray data, respectively.}
\end{table}
\begin{table}
\tabletypesize{\footnotesize}
\tablewidth{500pt}
\caption{Our results of joint spectral fits for the prompt gamma-rays (Slice 1) and afterglow (Slices 2-4) with a single power-law function. }
\centering
\begin{tabular}{cllc}
\hline\hline
  Slice  & Interval(s) & $\chi_r^{2}$ &  $\Gamma$   \\
\hline
  1   &    100-146&  $1.60$  &  $-1.32\pm0.01$   \\
  2   &  200-260& $1.27$   &  $-1.86\pm0.01$    \\
  3   &    690-800 & $0.96$   &   $-1.68\pm0.01$    \\
  4   &  27000-32000& $1.22$  &  $-1.57\pm0.01$    \\

\hline
\end{tabular}
\tablecomments{The hydrogen column density of Milky Way is fixed at $0.147\times10^{22} {\rm cm}^{-2}$. Optical extinction and neutral hydrogen absorbtion of soft X-rays of the GRB host galaxy are taken into account, but they are negligible.}
\end{table}

\begin{table}
\tiny
\tablewidth{450pt}
\caption{Results of our theoretical fits with the external shock models by considering the forward shock (marked with subscript ``f") and reverse shock (marked with subscript ``r") emission for GRB\,140512A. Fits to the reverse shock emission of GRBs 990123, 090102, and 130427A are also presented for comparison.}
\centering
\begin{tabular}{lllllllll}
\hline\hline
  GRB  &  $E_{\rm K,iso}$ &  $\Gamma_{0}$ &  $n$ &  $\theta_{\rm j}$ &  $\epsilon_{\rm e,\ f}$ &  $\epsilon_{\rm B,f}$ &$\epsilon_{\rm e,r}$ & $\epsilon_{\rm B,r}$  \\
  &  $(10^{54}\rm erg)$ &  &  $(\rm cm^{-3})$ &  (\rm rad) & &  ($\times10^{-8}$) & & ($\times10^{-4}$)  \\

\hline
  GRB\,140512A  & $(7.65\pm0.18)$&   $112.3\pm0.9$  &   $9.7\pm0.4$  &  $0.031\pm0.007$  &  $0.29\pm0.08$    &   $1.82^{+0.63}_{-0.84}$ &   $0.006\pm0.002$ &  $(1.49\pm0.06)$  \\
  GRB 990123$^{*}$    &$80$   &   $350$  &  $5$  &   $0.1$ & $0.05$  & $1.82$   & $0.02$  &  $1.49$    \\
  GRB 090102$^{*}$    &$4$   &   $180$  &  $18$  &   $-$  & $0.13$     & $1.82$  & $0.006$  & $1.49$    \\
  GRB 130427A$^{*}$  &$0.5$   &   $153$  &  $90$  &   $0.05$  & $0.31$    & $1.82$   & $0.006$  &  $1.49$    \\

\hline
\end{tabular}
\tablecomments{$^{*}$Fits to the reverse shock emission in GRBs\,990123, 090102, and 130427A are made by setting the $\epsilon_{\rm e,r}$ and $\epsilon_{\rm B,r}$ the same as that of GRB\,140512A for comparison. An much larger $\epsilon_{\rm e,r}$ is required to fit the reverse shock emission of GRB 990123 than that in other GRBs.}
\end{table}

\clearpage

\begin{thebibliography}{}
\bibitem[Akerlof et al.(1999)]{1999Natur.398..400A} Akerlof, C., Balsano, R., Barthelmy, S., et al.\ 1999, \nat, 398, 400


\bibitem[Chincarini et al.(2007)]{2007ApJ...671.1903C} Chincarini, G., Moretti, A., Romano, P., et al.\ 2007, \apj, 671, 1903


\bibitem[Colgate(1974)]{1974ApJ...187..333C} Colgate, S.~A.\ 1974, \apj, 187, 333


\bibitem[de Ugarte Postigo et al.(2014)]{2014GCN..16253...1D} de Ugarte Postigo, A., Gorosabel, J., Xu, D., et al.\ 2014a, GRB Coordinates Network, 16253, 1


\bibitem[de Ugarte Postigo et al.(2014)]{2014GCN..16310...1D} de Ugarte Postigo, A., Gorosabel, J., Xu, D., et al.\ 2014b, GRB Coordinates Network, 16310, 1


\bibitem[Eichler et al.(1989)]{1989Natur.340..126E} Eichler, D., Livio, M., Piran, T., \& Schramm, D.~N.\ 1989, \nat, 340, 126


\bibitem[Evans et al.(2010)]{2010A&A...519A.102E} Evans, P.~A., Willingale, R., Osborne, J.~P., et al.\ 2010, \aap, 519, A102


\bibitem[Fan et al.(2004)]{2004A&A...424..477F} Fan, Y.~Z., Wei, D.~M., \& Wang, C.~F.\ 2004, \aap, 424, 477


\bibitem[Fan et al.(2002)]{2002ChJAA...2..449F} Fan, Y.-Z., Dai, Z.-G., Huang, Y.-F., \& Lu, T.\ 2002, Chin. J. Astro. Astrophys., 2, 449


\bibitem[Fan \& Piran(2006)]{2006MNRAS.369..197F} Fan, Y., \& Piran, T.\ 2006, \mnras, 369, 197


\bibitem[Gao et al.(2013)]{2013NewAR..57..141G} Gao, H., Lei, W.-H., Zou, Y.-C., Wu, X.-F., \& Zhang, B.\ 2013, New Astronomy Reviews, 57, 141


\bibitem[Gao et al.(2015)]{2015ApJ...810..160G} Gao, H., Wang, X.-G., M{\'e}sz{\'a}ros, P., \& Zhang, B.\ 2015, \apj, 810, 160


\bibitem[Gehrels et al.(2004)]{2004ApJ...611.1005G} Gehrels, N., Chincarini, G., Giommi, P., et al.\ 2004, \apj, 611, 1005


\bibitem[Golenetskii et al.(2014)]{2014GCN..16265...1G} Golenetskii, S., Aptekar, R., Frederiks, D., et al.\ 2014, GRB Coordinates Network, 16265, 1


\bibitem[Gomboc et al.(2008)]{2008ApJ...687..443G} Gomboc, A., Kobayashi, S., Guidorzi, C., et al.\ 2008, \apj, 687, 443-455


\bibitem[Gomboc et al.(2009)]{2009AIPC.1133..145G} Gomboc, A., Kobayashi, S., Mundell, C.~G., et al.\ 2009, American Institute of Physics Conference Series, 1133, 145


\bibitem[Graham et al.(2014)]{2014GCN..16257...1G} Graham, J., Varela, K., Delvaux, C., \& Greiner, J.\ 2014, GRB Coordinates Network, 16257, 1


\bibitem[Harrison \& Kobayashi(2013)]{2013ApJ...772..101H} Harrison, R., \& Kobayashi, S.\ 2013, \apj, 772, 101


\bibitem[Huang et al.(2000)]{2000ApJ...543...90H} Huang, Y.~F., Gou, L.~J., Dai, Z.~G., \& Lu, T.\ 2000, \apj, 543, 90


\bibitem[Japelj et al.(2014)]{2014ApJ...785...84J} Japelj, J., Kopa{\v c}, D., Kobayashi, S., et al.\ 2014, \apj, 785, 84


\bibitem[Jin \& Fan(2007)]{2007MNRAS.378.1043J} Jin, Z.~P., \& Fan, Y.~Z.\ 2007, \mnras, 378, 1043


\bibitem[Kobayashi \& Zhang(2003)]{2003ApJ...597..455K} Kobayashi, S., \& Zhang, B.\ 2003, \apj, 597, 455


\bibitem[Kobayashi \& Zhang(2003)]{2003ApJ...582L..75K} Kobayashi, S., \& Zhang, B.\ 2003b, \apjl, 582, L75


\bibitem[Kopa{\v c} et al.(2013)]{2013ApJ...772...73K} Kopa{\v c}, D., Kobayashi, S., Gomboc, A., et al.\ 2013, \apj, 772, 73




\bibitem[Kumar \& Panaitescu(2003)]{2003MNRAS.346..905K} Kumar, P., \& Panaitescu, A.\ 2003, \mnras, 346, 905


\bibitem[Kumar \& Zhang(2015)]{2015PhR...561....1K} Kumar, P., \& Zhang, B.\ 2015, \physrep, 561, 1


\bibitem[L{\"u} \& Zhang(2014)]{2014ApJ...785...74L} L{\"u}, H.-J., \& Zhang, B.\ 2014, \apj, 785, 74


\bibitem[L{\"u} et al.(2015)]{2015ApJ...805...89L} L{\"u}, H.-J., Zhang, B., Lei, W.-H., Li, Y., \& Lasky, P.~D.\ 2015, \apj, 805, 89


\bibitem[Laskar et al.(2013)]{2013ApJ...776..119L} Laskar, T., Berger, E., Zauderer, B.~A., et al.\ 2013, \apj, 776, 119


\bibitem[Li et al.(2012)]{2012ApJ...758...27L} Li, L., Liang, E.-W., Tang, Q.-W., et al.\ 2012, \apj, 758, 27


\bibitem[Liang et al.(2009)]{2009ApJ...707..328L} Liang, E.-W., L{\"u}, H.-J., Hou, S.-J., Zhang, B.-B., \& Zhang, B.\ 2009, \apj, 707, 328


\bibitem[Liang et al.(2013)]{2013ApJ...774...13L} Liang, E.-W., Li, L., Gao, H., et al.\ 2013, \apj, 774, 13


\bibitem[Liang et al.(2008)]{2008ApJ...675..528L} Liang, E.-W., Racusin, J.~L., Zhang, B., Zhang, B.-B., \& Burrows, D.~N.\ 2008, \apj, 675, 528-552


\bibitem[Liang et al.(2010)]{2010ApJ...725.2209L} Liang, E.-W., Yi, S.-X., Zhang, J., et al.\ 2010, \apj, 725, 2209


\bibitem[Liang et al.(2006)]{2006ApJ...653L..81L} Liang, E.-W., Zhang, B.-B., Stamatikos, M., et al.\ 2006, \apjl, 653, L81


\bibitem[M{\'e}sz{\'a}ros \& Rees(1999)]{1999MNRAS.306L..39M} M{\'e}sz{\'a}ros, P., \& Rees, M.~J.\ 1999, \mnras, 306, L39


\bibitem[M{\'e}sz{\'a}ros \& Rees(1997)]{1997ApJ...476..232M} M{\'e}sz{\'a}ros, P., \& Rees, M.~J.\ 1997, \apj, 476, 232


\bibitem[MacFadyen \& Woosley(1999)]{1999ApJ...524..262M} MacFadyen, A.~I., \& Woosley, S.~E.\ 1999, \apj, 524, 262


\bibitem[Meegan et al.(2009)]{2009ApJ...702..791M} Meegan, C., Lichti, G., Bhat, P.~N., et al.\ 2009, \apj, 702, 791-804


\bibitem[Melandri et al.(2014)]{2014A&A...565A..72M} Melandri, A., Covino, S., Rogantini, D., et al.\ 2014, \aap, 565, A72


\bibitem[Melandri et al.(2010)]{2010ApJ...723.1331M} Melandri, A., Kobayashi, S., Mundell, C.~G., et al.\ 2010, \apj, 723, 1331


\bibitem[Melandri et al.(2008)]{2008ApJ...686.1209M} Melandri, A., Mundell, C.~G., Kobayashi, S., et al.\ 2008, \apj, 686, 1209-1230


\bibitem[Mundell et al.(2007)]{2007Sci...315.1822M} Mundell, C.~G., Steele, I.~A., Smith, R.~J., et al.\ 2007, Science, 315, 1822


\bibitem[Nakar \& Piran(2004)]{2004MNRAS.353..647N} Nakar, E., \& Piran, T.\ 2004, \mnras, 353, 647


\bibitem[Narayan et al.(1992)]{1992ApJ...395L..83N} Narayan, R., Paczynski, B., \& Piran, T.\ 1992, \apjl, 395, L83


\bibitem[Nousek et al.(2006)]{2006ApJ...642..389N} Nousek, J.~A., Kouveliotou, C., Grupe, D., et al.\ 2006, \apj, 642, 389


\bibitem[O'Brien et al.(2006)]{2006ApJ...647.1213O} O'Brien, P.~T., Willingale, R., Osborne, J., et al.\ 2006, \apj, 647, 1213


\bibitem[Oates et al.(2009)]{2009MNRAS.395..490O} Oates, S.~R., Page, M.~J., Schady, P., et al.\ 2009, \mnras, 395, 490


\bibitem[Paczynski(1986)]{1986ApJ...308L..43P} Paczynski, B.\ 1986, \apjl, 308, L43


\bibitem[Pagani et al.(2014)]{2014GCN..16249...1P} Pagani, C., Barthelmy, S.~D., Burrows, D.~N., et al.\ 2014, GRB Coordinates Network, 16249, 1


\bibitem[Panaitescu(2005)]{2005MNRAS.363.1409P} Panaitescu, A.\ 2005, \mnras, 363, 1409


\bibitem[Panaitescu \& Kumar(2001)]{2001ApJ...554..667P} Panaitescu, A., \& Kumar, P.\ 2001, \apj, 554, 667


\bibitem[Peng et al.(2014)]{2014ApJ...795..155P} Peng, F.-K., Liang, E.-W., Wang, X.-Y., et al.\ 2014, \apj, 795, 155


\bibitem[Qin et al.(2013)]{2013ApJ...763...15Q} Qin, Y., Liang, E.-W., Liang, Y.-F., et al.\ 2013, \apj, 763, 15


\bibitem[Resmi \& Zhang(2016)]{2016ApJ...825...48R} Resmi, L., \& Zhang, B.\ 2016, \apj, 825, 48


\bibitem[Roming et al.(2006)]{2006ApJ...652.1416R} Roming, P.~W.~A., Schady, P., Fox, D.~B., et al.\ 2006, \apj, 652, 1416


\bibitem[Rossi \& Rees(2003)]{2003MNRAS.339..881R} Rossi, E., \& Rees, M.~J.\ 2003, \mnras, 339, 881


\bibitem[Sakamoto et al.(2014)]{2014GCN..16258...1S} Sakamoto, T., Barthelmy, S.~D., Baumgartner, W.~H., et al.\ 2014, GRB Coordinates Network, 16258, 1


\bibitem[Sari \& Piran(1999)]{1999ApJ...520..641S} Sari, R., \& Piran, T.\ 1999a, \apj, 520, 641


\bibitem[Sari \& Piran(1999)]{1999ApJ...517L.109S} Sari, R., \& Piran, T.\ 1999b, \apjl, 517, L109


\bibitem[Sari et al.(1998)]{1998ApJ...497L..17S} Sari, R., Piran, T., \& Narayan, R.\ 1998, \apjl, 497, L17


\bibitem[Schlafly \& Finkbeiner(2011)]{2011ApJ...737..103S} Schlafly, E.~F., \& Finkbeiner, D.~P.\ 2011, \apj, 737, 103


\bibitem[Stanbro(2014)]{2014GCN..16262...1S} Stanbro, M.\ 2014, GRB Coordinates Network, 16262, 1


\bibitem[Steele et al.(2009)]{2009Natur.462..767S} Steele, I.~A., Mundell, C.~G., Smith, R.~J., Kobayashi, S., \& Guidorzi, C.\ 2009, \nat, 462, 767


\bibitem[Wang et al.(2013)]{2013ApJ...774..132W} Wang, X.-G., Liang, E.-W., Li, L., et al.\ 2013, \apj, 774, 132


\bibitem[Wang et al.(2015)]{2015ApJS..219....9W} Wang, X.-G., Zhang, B., Liang, E.-W., et al.\ 2015, \apjs, 219, 9


\bibitem[Willingale et al.(2013)]{2013MNRAS.431..394W} Willingale, R., Starling, R.~L.~C., Beardmore, A.~P., Tanvir, N.~R., \& O'Brien, P.~T.\ 2013, \mnras, 431, 394


\bibitem[Woosley(1993)]{1993ApJ...405..273W} Woosley, S.~E.\ 1993, \apj, 405, 273


\bibitem[Woosley \& Bloom(2006)]{2006ARA&A..44..507W} Woosley, S.~E., \& Bloom, J.~S.\ 2006, \araa, 44, 507


\bibitem[Wu et al.(2003)]{2003MNRAS.342.1131W} Wu, X.~F., Dai, Z.~G., Huang, Y.~F., \& Lu, T.\ 2003, \mnras, 342, 1131


\bibitem[Xin et al.(2014)]{2014GCN..16261...1X} Xin, L.~P., Wang, X.~F., Wei, J.~Y., et al.\ 2014, GRB Coordinates Network, 16261, 1


\bibitem[Xin et al.(2011)]{2011MNRAS.410...27X} Xin, L.-P., Liang, E.-W., Wei, J.-Y., et al.\ 2011, \mnras, 410, 27


\bibitem[Xin et al.(2016)]{2016ApJ...817..152X} Xin, L.-P., Wang, Y.-Z., Lin, T.-T., et al.\ 2016, \apj, 817, 152


\bibitem[Yi et al.(2013)]{2013ApJ...776..120Y} Yi, S.-X., Wu, X.-F., \& Dai, Z.-G.\ 2013, \apj, 776, 120


\bibitem[Yi et al.(2016)]{2016ApJS..224...20Y} Yi, S.-X., Xi, S.-Q., Yu, H., et al.\ 2016, \apjs, 224, 20


\bibitem[Yost et al.(2003)]{2003ApJ...597..459Y} Yost, S.~A., Harrison, F.~A., Sari, R., \& Frail, D.~A.\ 2003, \apj, 597, 459


\bibitem[Zaninoni et al.(2013)]{2013A&A...557A..12Z} Zaninoni, E., Bernardini, M.~G., Margutti, R., Oates, S., \& Chincarini, G.\ 2013, \aap, 557, A12


\bibitem[Zhang(2014)]{2014IJMPD..2330002Z} Zhang, B.\ 2014, International Journal of Modern Physics D, 23, 1430002


\bibitem[Zhang et al.(2006)]{2006ApJ...642..354Z} Zhang, B., Fan, Y.~Z., Dyks, J., et al.\ 2006, \apj, 642, 354


\bibitem[Zhang \& Kobayashi(2005)]{2005ApJ...628..315Z} Zhang, B., \& Kobayashi, S.\ 2005, \apj, 628, 315


\bibitem[Zhang et al.(2003)]{2003ApJ...595..950Z} Zhang, B., Kobayashi, S., \& M{\'e}sz{\'a}ros, P.\ 2003a, \apj, 595, 950


\bibitem[Zhang et al.(2007)]{2007RSPTA.365.1257Z} Zhang, B., Liang, E., Gupta, N., et al.\ 2007, Philosophical Transactions of the Royal Society of London Series A, 365, 1257


\bibitem[Zhang \& Yan(2011)]{2011ApJ...726...90Z} Zhang, B., \& Yan, H.\ 2011, \apj, 726, 90

\bibitem[Zhang et al.(2015)]{2015ApJ...798....3Z} Zhang, S., Jin, Z.-P., \& Wei, D.-M.\ 2015, \apj, 798, 3


\bibitem[Zhang et al.(2003)]{2003ApJ...586..356Z} Zhang, W., Woosley, S.~E., \& MacFadyen, A.~I.\ 2003b, \apj, 586, 356


\bibitem[Zou et al.(2005)]{2005MNRAS.363...93Z} Zou, Y.~C., Wu, X.~F., \& Dai, Z.~G.\ 2005, \mnras, 363, 93



\end{thebibliography}
\end{document}